# Micromechanics analysis of thermal expansion and thermal pressurization of a hardened cement paste

Siavash Ghabezloo[*]

*Université Paris-Est, Laboratoire Navier, CERMES, Ecole des Ponts ParisTech, Marne la Vallée, France*

## Abstract

The results of a macro-scale experimental study of the effect of heating on a fluid-saturated hardened cement paste are analysed using a multi-scale homogenization model. The analysis of the experimental results revealed that the thermal expansion coefficient of the cement paste pore fluid is anomalously higher than the one of pure bulk water. The micromechanics model is calibrated using the results of drained and undrained heating tests and permits the extrapolation of the experimentally evaluated thermal expansion and thermal pressurization parameters to cement pastes with different water-to-cement ratios. It permits also to calculate the pore volume thermal expansion coefficient $\alpha_\phi$ which is difficult to evaluate experimentally. The anomalous pore fluid thermal expansion is also analysed using the micromechanics model.

---

[*] Siavash Ghabezloo, CERMES, Ecole des Ponts ParisTech, 6-8 avenue Blaise Pascal, Cité Descartes, 77455 Champs-sur-Marne, Marne la Vallée cedex 2, France. Email: siavash.ghabezloo@enpc.fr

# 1. Introduction

A better understanding of the effects of undrained heating and induced thermal pressurization phenomenon is an important point to properly understand the behaviour and evaluate the integrity of an oil well cement sheath submitted to rapid temperature changes. Temperature increase in saturated porous materials under undrained conditions leads to volume change and pore fluid pressure increase. This thermal pressurization is due to discrepancy between thermal expansion coefficients of the pore fluid and of the pore volume. This pore pressure increase induces a reduction of effective mean stress and can lead to shear failure or hydraulic fracturing. Indeed the geomaterials are pressure sensitive and the maximum shear stress depends on the effective mean stress. On the other hand, when pore pressure is higher than maximum principal stress (positive in compression) the material may undergo tensile failure. The thermal pressurization phenomenon is important in petroleum engineering where the reservoir rock and the well cement lining undergo sudden temperature changes, as for example when extracting heavy oils by steam injection methods. This rapid temperature increase could damage cement sheath integrity of wells and lead to loss of zonal isolation. Within this context, a macro-scale experimental program of drained and undrained heating tests is performed on a fluid-saturated hardened oil-well cement paste. The results of this study are presented in [1] and show some important aspects of behaviour of this material when submitted to rapid temperature changes. The thermal pressurization coefficient, defined as the pore pressure increase due to a unit temperature increase in undrained conditions, is evaluated to about 0.6MPa/°C which is a relatively high value comparing to other geomaterials (see [2] for a review). In accordance with results of Valenza and Scherer [3], the analysis of the undrained heating test revealed that the thermal expansion coefficient of cement paste pore fluid is anomalously higher than the one of pure bulk water. This experimental study was a part of a larger study on the thermo-poro-mechanical behaviour of a hardened cement paste [1][4][5][6]. The experimental program consisted by drained, undrained and unjacketed compression tests, as well as drained and undrained heating tests and permeability evaluation tests. This experimental program is performed on a particular cement paste, prepared with class G cement at a water-to-cement ratio equal to 0.44. The poroelastic parameters are then extrapolated to cement pastes with different water-to-cement ratios by means of micromechanics modelling and homogenization technique [7]. This is done using a multi-scale micromechanics model, originally proposed by Ulm *et al*. [8], which is calibrated on the experimental results [7]. The predictive capacity of the micromechanics model is verified by comparing the parameters with some experimental results from literature. In the continuity of the approach used in [7], the micromechanics model is used here in association with the results of drained and undrained heating tests presented in [1]. The model has been already calibrated in [7] for the poroelastic parameters, but a second calibration step should be performed for the thermal behaviour. This permits also to study the thermal expansion of the pore fluid in different parts of the microstructure. The calibrated model will then be used to calculate thermal expansion and thermal pressurization parameters for cement pastes with different water-to-cement ratios.

The paper is organized in six sections and one appendix. After this introduction, the second section presents the microstructure of the cement paste and discusses the anomalous thermal expansion of its pore fluid. The third section presents the theoretical framework of macro- and micro-thermo-poroelasticity and homogenization method. The results and analysis of drained and undrained heating tests are presented briefly in the fourth section. The homogenization of thermal expansion and thermal pressurization properties is presented in the fifth section. First the calibration of the model is completed on the basis of the results of drained and undrained heating tests and then, the homogenization model is used to extrapolate these experimental results for the cement pastes with different water-to-cement ratios. The last section is dedicated to concluding remarks. Appendix A presents the developments of micro-thermo-poroelasticity equations.

## 2. Microstructure of cement paste

The cement clinker is composed of four main phases: $C_3S$, $C_2S$, $C_3A$ and $C_4AF$ where in the standard cement chemistry the notation C stands for CaO, S for $SiO_2$, A for $Al_2O_3$ and F for $Fe_2O_3$. The setting and hardening of cement paste are the results of complex reactions, called hydration reactions, between clinker phases and water. The cement paste has a very complex microstructure which varies with cement composition, time and hydration conditions. In a simplified view, the main phases of the microstructure are calcium-silicate-hydrate (C−S−H) which is the main binding phase of all Portland cement-based materials, Portlandite (CH), Aluminates (AL), cement clinkers (CK) and macro-porosity. C−S−H is the main hydration product which is a porous phase with an amorphous and colloidal structure and a variable chemical composition. The CH often occurs as massive crystals but is also mixed with C−S−H at the micron-scale. CH and cement clinker can be considered as non-porous solid phases. Because of its colloidal and amorphous nature and the variability of its chemical composition, the structure of C−S−H matrix and its solid phase are not clearly known. Since a few decades different models have been proposed in the literature for the structure of this material. Most of these models consider a layered structure for C−S−H and also the existence of an important quantity of chemically bonded or adsorbed water. Jennings [9][10] proposed a microstructural model for C−S−H in which the amorphous and colloidal structure of the C−S−H is organized in elements, called 'globules'. The globule, with a size of about 4nm, is composed of solid C−S−H sheets, intra-globule porosity filled with structural water and a monolayer of water on the surface. The structure of C−S−H in Jennings' model contains small gel pores in the space between adjacent globules and larger gel pores between the groups of several globules. Jennings' model distinguishes two types of C−S−H, called low density (LD) and high density (HD) C−S−H. The globules are considered to be identical in LD and HD C−S−H and the difference between these two types of C−S−H is in the gel porosity of about 0.24 for HD C−S−H and 0.37 for LD C−S−H. A more detailed description of the microstructure of the hardened cement paste is presented in [7]. For the purpose of micromechanics modelling the microstructure of the cement paste is divided into the following three scale levels:

- Level 0 ($10^{-9}$–$10^{-8}$m, the C–S–H solid): Solid phase of C–S–H matrix.
- Level 1 ($10^{-8}$–$10^{-6}$m, the C–S–H matrix): High density and low density C–S–H.
- Level 2 ($10^{-6}$–$10^{-4}$m, the cement paste): C–S–H matrix, Portlandite (CH), Alluminates (AL), cement clinker (CK) and water.

## 2.1. Active porosity

The analysis of the test results of Ghabezloo *et al*. [4] revealed that the active porosity of the cement paste in poromechanics tests is smaller than its total porosity. A pore volume can be considered to be active if under the effect of a pressure gradient, the pore fluid can exchange with the fluid filling the pore volume situated in its neighbourhood. From the poromechanics point of view, the inactive pore volume and the pore fluid filling it should be considered as a part of the solid phase. Scherer *et al*. [11] and Sun and Scherer [12][13] also argued that a part of the pore fluid in the microstructure of cement paste is inactive. Accordingly, these authors reduced the total porosity of cement paste and mortar samples for the effect of inactive porosity.

The distribution of the active pore volume within the total pore volume of cement paste is not accurately known, but is important for homogenization of the poroelastic properties. Considering the microstructure of the hardened cement paste, it seems reasonable to assume that the inactive porosity is entirely situated in HD C–S–H. A detailed discussion of this assumption is presented in [7]. Consequently, the active porosity consists of the porosity in LD C–S–H and the macro-porosity. Assuming that the porosity in HD C–S–H is not active from the poromechanics point of view means that in the time-scale of the applied loads, the mass of the pore fluid in HD C–S–H porosity is constant. Consequently the HD C–S–H behaves like a porous material in undrained conditions.

## 2.2. Anomalous pore fluid thermal expansion

There are experimental evidences showing that the thermal expansion coefficient of cement paste pore fluid is higher than the one of pure bulk water [3][1]. This phenomenon is mainly attributed to confinement of pore fluid in very small pores of the microstructure. It is known that the thermal expansion coefficient of fluids when confined in very small pores, smaller than 15nm, is anomalously higher than that of the bulk fluid. This is confirmed by experimental evaluations of thermal expansion of water and salt solutions in porous silica glasses [14][15][16][17] showing that the thermal expansion of confined fluid increases with decreasing pore size. Moreover, the ratio of the thermal expansion of confined fluid to that of bulk fluid decreases with temperature increase.

The origin of this anomalous thermal expansion is not clearly known but is attributed to surface effects resulting in higher pressure of the fluid in the close vicinity of solid surface [18] or the disturbance of the structure of water molecules in a thin layer adjacent to the solid surface [19]. Considering the pressure dependency of thermal expansion of water, the higher pore pressure at the vicinity of the pore wall results in an average thermal expansion of fluid in a small pore that is different from the one of bulk fluid. Figure (1) presents the thermal expansion of pure bulk water as a function of pressure at various temperatures [20]. We can see that for temperatures below 50°C the thermal expansion increases with pressure, while above 50°C it decreases with pressure increase. At 50°C the thermal expansion of water shows almost no pressure dependency. From these observations, the average thermal expansion of water in a pore for temperatures below 50°C is higher than the one of bulk water, and the thermal expansion anomaly decreases with pore size and temperature increase. This analysis for temperature below 50°C is therefore compatible with the experimental results on the behaviour of confined fluids mentioned above. A similar compatibility can also be observed for the results of the analysis performed by Garofalini *et al*. [19].

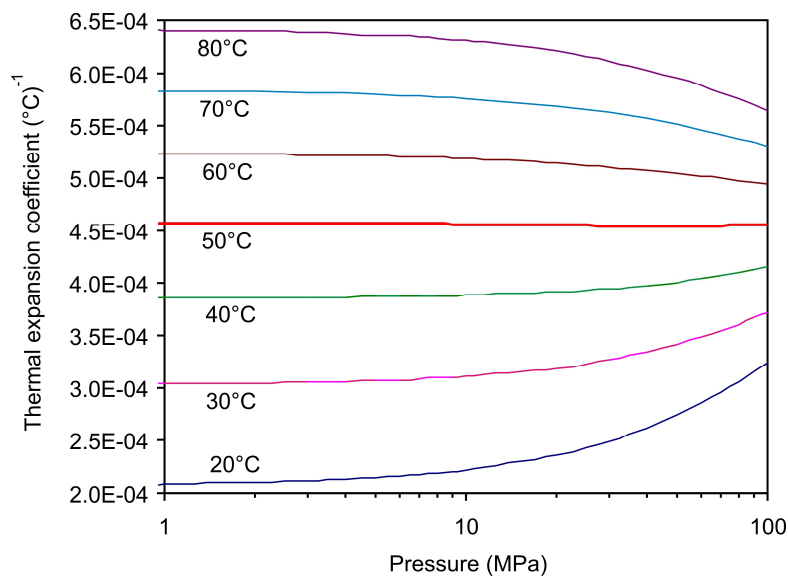

**Figure (1): Pressure dependency of thermal expansion of pure bulk water at different temperatures**

Valenza and Scherer [3] were the first who noticed the anomalous thermal expansion of cement paste pore fluid when comparing permeability measurements using two different methods: thermopermeametry and beam bending. According to these authors, to bring the two measurements into agreement, the pore fluid should have a thermal expansion coefficient about one and a half times larger than the one of bulk fluid. This is confirmed by experimental study of Ghabezloo *et al*. [1], presented in section 4, who showed that the pore fluid thermal expansion is greater than the one of pure bulk water and has a lower rate of increase with temperature.

In addition to confinement in small pores, it seems that the presence of dissolved ions in the pore fluid of cement paste influences its thermal expansion. It is known that the presence of dissolved ions in a fluid increases its thermal expansion in both bulk and confined conditions [21][17].

Typical concentrations in the pore fluid of a high alkali cement paste for $Na^+$, $K^+$ and $OH^-$ are respectively 0.16mol/l, 0.55mol/l and 0.71mol/l after 180 days [22]. Accordingly Ghabezloo *et al.* [1] argued that the anomalous thermal expansion of cement paste pore fluid is partly due to the presence of dissolved ions in the pore fluid.

## *2.3. Volume fractions*

The volume fractions of different phases of the microstructure of cement paste can be evaluated by knowing cement composition, water-to-cement ratio and degree of hydration. Using the method proposed by Bernard *et al.* [23], presented in details in [7], one can evaluate the volume fractions of C-S-H, Portlandite, unhydrated clinker and macro-porosity. These volume fractions are denoted respectively by $f_{CSH}$, $f_{CH}$, $f_{CK}$ and $f_V$. The volume fractions that are calculated by assuming the complete hydration are presented in Table (1). The parameter $f_{CSH}^{LD}$ gives the volume fractions of HD and LD C-S-H as $f_{LD} = f_{CSH}^{LD} f_{CSH}$ and $f_{HD} = \left(1 - f_{CSH}^{LD}\right) f_{CSH}$. It is assumed that the Aluminates phase has the same properties as the C−S−H phase [7], consequently the volume fraction of C−S−H in Table (1) is equal to the sum of volume fractions of Aluminates and C−S−H. The volume fractions presented in Table (1) are used in a micromechanics model to evaluate the macroscopic properties of the hardened cement paste and to analyse the results of macro-scale drained and undrained heating tests. The needed theoretical framework for doing this analysis is presented in the following section.

| C−S−H level | | Pore fluid | | Cement paste level | |
|---|---|---|---|---|---|
| **Parameter** | **Value** | **Parameter** | **Value** | **Parameter** | **Value** |
| $k_s$ | Calibrated (25.0 GPa)* | $\alpha_{f,HD}$ | calibrated | $f_{CSH}$ | 0.71 |
| $g_s$ | Calibrated (18.4 GPa)* | $\alpha_{f,LDV}$ | calibrated | $f_{CH}$ | 0.18 |
| $\alpha_s$ | calibrated | | | $f_V$ | 0.11 |
| $\phi_{HD}$ | 0.24 | | | $f_{CK}$ | 0.00 |
| $\phi_{LD}$ | 0.37 | | | $f_{CSH}^{LD}$ | 0.60 |
| | | | | $k_{CH}$ | 32.5 GPa |
| | | | | $g_{CH}$ | 14.6 GPa |
| | | | | $\alpha_{CH}$ | $7 \times 10^{-5}$ (°C)$^{-1}$ |

* Calibration presented in [7]

**Table (1): Homogenization model parameters**

# 3. Theoretical framework

This paper associates the results of a macro-scale experimental study with the tools of micro-mechanics theory and homogenization method. It is therefore necessary to present the theoretical framework used in the macro-scale experimental study, the one used for the micromechanics modelling, as well as the link between the parameters in these two scales. The theoretical framework is the same as the one presented in [7], but is extended here to take into account the temperature effect.

## *3.1. Macro-thermo-poroelasticity*

The theoretical framework is presented for the macroscopic thermo-elastic volumetric behaviour of a porous material which is heterogeneous at the micro-scale. This framework is presented in different papers and textbooks, e.g. [24][25][26][27][28], and is recalled briefly in the following. More details about the definitions of the parameters introduced in this section and the relations between these parameters can be found in [1][4][7].

The variations of the total volume $V$ and of the pore volume $V_\phi$ of an elementary volume introduce six parameters:

$$-\frac{dV}{V_0} = \frac{1}{K_d}d\Sigma_d + \frac{1}{K_s}dp - \alpha_d dT \qquad (1)$$

$$-\frac{dV_\phi}{V_{\phi 0}} = \frac{1}{K_p}d\Sigma_d + \frac{1}{K_\phi}dp - \alpha_\phi dT \qquad (2)$$

where $p$ is the pore pressure, $T$ is the temperature, $\Sigma = 1/3\mathbf{\Sigma}:\mathbf{1}$ is the isotropic stress which is positive in compression and $\Sigma_d = \Sigma - p$ is the differential pressure that is equivalent to Terzaghi effective stress. The macroscopic volumetric strain increment is defined as $dE = -dV/V_0$ ($E = \mathbf{E}:\mathbf{1}$). $K_d$ is the drained bulk modulus and $K_s$ is the unjacketed modulus. $K_p$ and $K_\phi$ are two moduli related to the pore volume. Using Betti's reciprocal theorem it can be shown that $\phi_0/K_p = 1/K_d - 1/K_s$ [29]. $\alpha_d$ is the volumetric drained thermal expansion coefficient that can be measured in a drained heating test and $\alpha_\phi$ is the pore volume thermal expansion coefficient. Like for $K_\phi$, the direct experimental evaluation of $\alpha_\phi$ is very difficult [1]. In the case of a micro-homogeneous and micro-isotropic porous material $\alpha_d = \alpha_\phi = \alpha_m$ and $K_s = K_\phi = K_m$, where $\alpha_m$ and $K_m$ are respectively thermal expansion coefficient and bulk modulus of the single solid constituent of the material.

The macro-scale experimental study of thermo-poromechanical behaviour of the hardened cement paste in [1][4][5][6] was based on the constitutive laws (1) and (2). The derivation of the equations of micro-thermo-poroelasticity, presented in section 3.2 and in Appendix A, is done more commonly based on an alternative set of parameters. These alternative parameters can be obtained by writing the variations of Lagrangian porosity and total stress from equations (1) and (2). The variation of Lagrangian porosity $\phi = V_\phi / V_0$ is given by:

$$d\phi = -bdE + \frac{1}{N}dp - QdT \quad ; \quad Q = b\alpha_d - \phi_0 \alpha_\phi \tag{3}$$

where $b$ is Biot's effective stress coefficient and $N$ is Biot's skeleton modulus. The parameter $Q$ gives the variations of the porosity with temperature when strains and pore pressure are constant. The total stress increment is given by:

$$d\Sigma = K_d dE + bdp + \kappa dT \quad ; \quad \kappa = K_d \alpha_d \tag{4}$$

The parameter $\kappa$ gives the variations of the total stress with temperature when the strains and the pore pressure are constant. The equations of micro-thermo-poroelasticity in the following section are presented in terms of $K_d$, $b$, $N$, $\kappa$ and $Q$.

An important part of the experimental study in [1][4] was based on undrained compression and heating tests. The analysis of the performed undrained heating test needs the introduction of the parameters $\Lambda$ and $\alpha_u$ that are measured in this test. The variation of the fluid content $m_f = \phi_0 \rho_f$ in the undrained conditions is equal to zero ($dm_f = 0$). The pore pressure variation in the undrained conditions is given by:

$$dp = Bd\Sigma + \Lambda dT \quad ; \quad \Lambda = \frac{Q + \phi_0 \alpha_f - b\alpha_d}{\frac{1}{N} + \frac{\phi_0}{K_f} + \frac{b^2}{K_d}} \tag{5}$$

where $\Lambda$ is the thermal pressurization coefficient and $B$ is Skempton's coefficient. $\alpha_f$ and $K_f$ are respectively the fluid thermal expansion coefficient and compression modulus. The volumetric strain in undrained conditions is given by:

$$dE = \frac{1}{K_u} d\Sigma - \alpha_u dT \quad ; \quad \alpha_u = \alpha_d + \frac{b\Lambda}{K_d} \tag{6}$$

where $\alpha_u$ is the undrained thermal expansion coefficient and $K_u$ is the undrained bulk modulus. In laboratory experiments, the most commonly performed tests are drained, undrained and unjacketed compression tests as well as drained and undrained heating test which yield $K_d$, $K_p$, $K_u$, $B$, $K_s$, $\alpha_d$, $\alpha_u$ and $\Lambda$. On the other hand, the homogenization of thermo-poro-elastic parameters is more commonly done using $K_d$, $b$, $N$, $\kappa$ and $Q$. The presentation of the complete set of parameters permits to establish the link between the parameters that are easier to evaluate experimentally and the ones used more commonly in micromechanics theory.

## 3.2. Micro-thermo-poroelasticity and homogenization method

The aim of classical homogenization method is to replace an actual heterogeneous complex body by a fictitious homogeneous one that behaves globally in the same way. The theoretical framework of micromechanics modelling and homogenization method has been presented in different papers and textbooks, e.g. [8][30][31][32][33][34][35]. The principles and main equations of this framework are presented briefly in the following. A detailed derivation of homogenization equations is presented in Appendix A.

The volume $V_0$ of the REV of a heterogeneous material is composed of $n$ different phases with volumes $V_r$, $r = 1...n$, and volume fractions $f_r = V_r/V_0$. We consider that there is only one porous phase with volume $V_\phi$ and porosity $\phi = V_\phi/V_0$. The number of solid phases is therefore $m = n-1$ with total volume $V_s$. The tensors of elastic moduli and thermal expansion coefficients of each phase are denoted respectively by $\mathbf{c}_r$ and $\boldsymbol{\alpha}_r$.

The equations of micro-thermo-poroelasticity and homogenization of thermo-poroelastic properties can be derived on a REV submitted to a homogeneous strain boundary condition and two eigenstresses, as presented in appendix A. The tensor of the effective elastic moduli $\mathbf{C}^{\text{hom}}$ is given by:

$$\mathbf{C}^{\text{hom}} = \langle \mathbf{c} : \mathbf{A} \rangle_V = \sum_{r=1}^{n} f_r \mathbf{c}_r : \langle \mathbf{A} \rangle_{V_r} \tag{7}$$

Where $\langle \mathbf{A} \rangle_{V_r}$ is the volume average of the strain localization tensors over the phase $r$ (see appendix A). The tensor of effective Biot's coefficients $\mathbf{b}$ is expressed as:

$$\mathbf{b}^{\text{hom}} = \phi_0 \mathbf{1} : \langle \mathbf{A} \rangle_{V_\phi} = \mathbf{1} : \left( \mathbf{I} - \sum_{r=1}^{m} f_r \langle \mathbf{A} \rangle_{V_r} \right) \tag{8}$$

The tensor of effective solid moduli $\mathbf{C}_s^{\text{hom}}$ can be identified by writing the relation between the average local stresses and strains over the solid volume:

$$\langle \boldsymbol{\sigma}' \rangle_{V_s} = \mathbf{C}_s^{\text{hom}} : \langle \boldsymbol{\varepsilon}' \rangle_{V_s} \quad ; \quad \mathbf{C}_s^{\text{hom}} = \langle \mathbf{c} : \mathbf{A} \rangle_{V_s} : \langle \mathbf{A} \rangle_{V_s}^{-1} \tag{9}$$

$\langle \mathbf{A} \rangle_{V_\phi}$ and $\langle \mathbf{A} \rangle_{V_s}$ are the volume average of the strain localization tensors respectively over the pore volume and the solid volume. The effective Biot's skeleton modulus $N^{\text{hom}}$ is given by:

$$\frac{1}{N^{\text{hom}}} = \mathbf{1} : \sum_{r=1}^{m} f_r \mathbf{c}_r^{-1} : \left( \mathbf{1} - \mathbf{1} : \langle \mathbf{A} \rangle_{V_r} \right) \tag{10}$$

The effective thermal parameter $\boldsymbol{\kappa}^{\text{hom}}$ and the effective thermal expansion $\boldsymbol{\alpha}_d^{\text{hom}}$ are given by:

$$\boldsymbol{\kappa}^{\text{hom}} = \langle \mathbf{c} : \boldsymbol{\alpha} : \mathbf{A} \rangle_V \quad ; \quad \boldsymbol{\alpha}_d^{\text{hom}} = \left( \mathbf{C}^{\text{hom}} \right)^{-1} : \boldsymbol{\kappa}^{\text{hom}} \tag{11}$$

The effective thermal parameter $Q^{\text{hom}}$ is obtained as:

$$Q^{\text{hom}} = \mathbf{1} : \sum_{r=1}^{m} f_r \boldsymbol{\alpha}_r : \left( \mathbf{I} - \langle \mathbf{A} \rangle_{V_r} \right) \tag{12}$$

### 3.2.1. Multi-scale porous material

Considering the multi-scale microstructure of the hardened cement paste presented in section 2, the homogenization of the macroscopic properties should be performed in two steps. The pore volume of the hardened cement paste manifests itself at two different scales. The scale I corresponds to the gel porosity of HD and LD C-S-H, and the scale II corresponds to the macro-porosity. The homogenization of the macroscopic properties of such materials can be done using a multi-step homogenization method [8][35]. This method is explained briefly in the following for a two-scale porous material in which the pore volume exhibits two different scales I and II ($V_\phi = V_\phi^I + V_\phi^{II}$). This material is composed of $l$ ($l \leq m$) porous phases with the porosities $\phi_r^I$, $m-l$ solid phases and a pore volume with the porosity $\phi^{II}$. The pore volumes are connected and there is one homogeneous pore pressure. The total porosity is given by:

$$\phi = \sum_{r=1}^{l} f_r \phi_r^I + \phi^{II} \tag{13}$$

At the level I of the microstructure of the cement paste LD and HD C-S-H are composed of one solid phase and one porous phase with nanometer size pores. At the level II the cement paste is composed of LD and HD C-S-H, Portlandite, unhydrated clinker and macro-porosity. Standard homogenization equations, as presented in the previous section, give the poroelastic properties of the $l$ porous phases of level I ($\mathbf{c}_r^I$, $\mathbf{b}_r^I$, $N_r^I$, $\mathbf{c}_{sr}^I$, $\mathbf{\alpha}_r^I$, $Q_r^I$). The effective parameters $\mathbf{C}^{hom}$ and $\mathbf{\kappa}^{hom}$ can be evaluated respectively from equations (7) and (11). The other effective parameters $b^{hom}$, $N^{hom}$ and $Q^{hom}$ for level II are given in the following equations.

$$\mathbf{b}^{hom} = \mathbf{1} - \sum_{r=1}^{m} \left( f_r \langle \mathbf{A} \rangle_{V_r} : \left( \mathbf{1} - \mathbf{b}_r^I \right) \right) \tag{14}$$

$$\frac{1}{N^{hom}} = \sum_{r=1}^{m} f_r \left( \mathbf{1} : \left( \mathbf{c}_{sr}^I \right)^{-1} : \left( \mathbf{I} - \langle \mathbf{A} \rangle_{V_r} \right) : \left( \mathbf{1} - \mathbf{b}_r^I \right) + \frac{1}{N_r^I} \right) \tag{15}$$

$$Q^{hom} = \sum_{r=1}^{m} f_r \left( \mathbf{\alpha}_r^I : \left( \mathbf{I} - \langle \mathbf{A} \rangle_{V_r} \right) : \left( \mathbf{1} - \mathbf{b}_r^I \right) + Q_r^I \right) \tag{16}$$

A detailed derivation of these equations is presented in Appendix A. It can be easily verified that when all solid phases are non-porous ($\mathbf{b}_r^I = 0$, $1/N_r^I = 0$, $\mathbf{c}_{sr}^I = \mathbf{c}_r$, $Q_r^I = 0$), equations (14), (15) and (16) reduce respectively to equations (8), (10) and (12).

## 4. Experimental evaluation of thermal expansion and thermal pressurization parameters

The experimental program for evaluation of the poroelastic parameters at ambient temperature is presented in [4][6] and is briefly recalled in [7]. To study the effect of temperature, drained and undrained heating tests have been performed under constant confining pressure. The results are presented in details in [1] and are briefly recalled in the following.

The tests were performed on cylindrical samples with 38mm diameter and 76mm length, made from class G oil well cement at w/c=0.44. The samples were cured for at least three months in a bath containing an equilibrated fluid under a controlled temperature of 90°C. This temperature was chosen to reproduce the curing conditions of a cement lining installed in a deep (~2 km) oil well. The porosity of the samples was studied by two methods: oven drying and mercury intrusion porosimetry. The total porosity was measured equal to $\phi = 0.35$ by drying the samples at 105°C until a constant mass is achieved. Mercury intrusion porosimetry was performed with a maximum intruding pressure of 200MPa and the porosity is equal to $\phi = 0.26$. The maximum intruding pressure of 200MPa corresponds to a minimum pore diameter of about 6nm.

A drained heating test was performed under a constant confining pressure of 1.5 MPa and a constant pore pressure of 1.0 MPa. During the test, the temperature was increased from 18 °C to 87 °C at a rate of 0.08 °C/min. The volumetric strain-temperature response is almost linear and results in a drained thermal expansion coefficient with negligible temperature dependency, equal to $6 \times 10^{-5} \left(°C\right)^{-1}$. The thermal pressurization phenomenon was studied in an undrained heating test under a constant confining pressure. In the beginning of the test the confining pressure was increased up to 19 MPa in drained conditions. After the stabilization of creep strains, the temperature was increased at a rate equal to 0.1 °C/min in undrained conditions and the pore pressure changes were monitored. The heating phase was continued until a point where the pore pressure reached the confining pressure at about 60°C. The heating of the sample was stopped at this point and the cooling phase was started. The measured pore pressure was corrected for the effect of the dead volume and thermo-mechanical deformations of drainage system of the triaxial cell using a simple method presented in [2][39]. The average thermal pressurization coefficient $\Lambda$ is found equal to 0.62 MPa/°C for heating phase and 0.57 MPa/°C for cooling phases. The analysis of test results showed that the variations of thermal pressurization coefficient is less that what is expected from the variations of thermal expansion of water with temperature. Moreover, the value of coefficient $\Lambda$ particularly at lower temperatures is higher than what can be evaluated by knowing other thermo-poro-elastic parameters. This unexpected thermal pressurization response is attributed to the anomalous thermal expansion behaviour of the cement paste pore fluid. This anomalous behaviour is discussed in details in section 2.2. The back analysis of pore pressure and volumetric strain responses of the undrained heating test permitted to evaluate the thermal expansion coefficient of pore fluid. The analysis showed that for temperatures between 25 and 55°C the pore fluid thermal expansion is greater than the one of pure bulk water and has a lower rate of increase with temperature. The analysis is performed on the cooling phase of the test, because of the less importance of the creep strains. For the same reason, the calibration of the micromechanics model in section 5.2 is done also on the cooling phase. For the sake of simplicity, the corrected pore pressure-temperature curve of the cooling phase is approximated by a hyperbolic equation, which gives the following linear expression for the measured thermal pressurization coefficient as a function of the temperature:

$$\Lambda\left(\text{MPa}/°C\right) = 0.4 + 0.0046T \qquad 25°C \leq T \leq 55°C \qquad (17)$$

This equation is used in section 5.2 for calibration of the homogenization model.

# 5. Homogenization of thermal expansion and thermal pressurization properties

The classical homogenization procedure is to calculate the homogenized macroscopic properties by knowing the microscopic parameters. In the method used in [7] and in this work, the macroscopic properties are already known for a particular cement paste and will be used to calibrate the unknown microscopic parameters, thermo-elastic parameters of C-S-H solid and thermal expansion of pore fluid in HD and LD C-S-H. The model will then permit to calculate the same macroscopic parameters for a cement paste having different water-to-cement ratio. A part of this procedure concerning the mechanical properties has been already done in [7].

## *5.1. Homogenization equations*

Considering the multi-scale microstructure of the cement paste, the homogenization of the macroscopic parameters is done in two steps, first for HD and LD C–S–H and then for the cement paste. A micromechanics representation of the REVs of C–S–H and cement paste which are used in the homogenization model is given in [7].

### 5.1.1. Level 1: C–S–H matrix

HD and LD C-S-H are constituted by one solid phase and one porous phase. The homogenization of poroelastic properties of C–S–H matrix can therefore be done using equations presented in section 3.2. The only difference between HD and LD C–S–H is in their packing density or porosity. The needed parameters are the porosities, $\phi_{LD}$ and $\phi_{HD}$ and the elastic parameters and the thermal expansion of the C–S–H solid, $k_s$, $g_s$ and $\alpha_s$. The homogenized drained bulk modulus and the shear modulus can be evaluated from equation (7) considering one solid phase and one porous phase. The subscript X represents LD or HD.

$$K_X^{hom} = (1-\phi_X)k_s A_{s,X}^v \quad ; \quad G_X^{hom} = (1-\phi_X)g_s A_{s,X}^d \tag{18}$$

Assuming an Eshelbian type morphology with spherical shapes for solids and pores, the strain localization tensor parameters, $A_{r,X}^v$ and $A_{r,X}^d$, of each phase can be estimated using equation (A.10) considering a solid and a porous phase, $(r = s, \phi)$. Note that this homogenization approach only uses the volume fractions of different phases for estimation of the thermo-poro-elastic properties and the grains size and the pore size distribution are not taken into account. Considering the self-consistent scheme we should take $k_0 = K_X^{hom}$, $g_0 = G_X^{hom}$ and the parameters $\alpha_0$ and $\beta_0$ are given by following relations:

$$\alpha_0 = \frac{3K_X^{hom}}{3K_X^{hom} + 4G_X^{hom}} \quad ; \quad \beta_0 = \frac{6(K_X^{hom} + 2G_X^{hom})}{5(3K_X^{hom} + 4G_X^{hom})} \tag{19}$$

Considering these expressions, the evaluation of $K_X^{hom}$ and $G_X^{hom}$ from equation (18) should be done using iterative calculations. The homogenized thermal parameters can be calculated from equations (11) and (12):

$$\kappa_X^{hom} = (1-\phi_X) k_s \alpha_s A_{s,X}^v \quad ; \quad \alpha_{d,X}^{hom} = \alpha_s \tag{20}$$

$$Q_X^{hom} = \alpha_s (1-\phi_X)(1-A_{s,X}^v) \tag{21}$$

The expressions of the other homogenized poroelastic parameters $b_X^{hom}$ and $N_X^{hom}$, presented in [7], can be obtained respectively from equations (8) and (10).

As mentioned in section 2.1, we assume that the porosity in HD C−S−H is not active so that the HD C−S−H behaves like a porous material in the undrained conditions. Consequently, in the second homogenization step the undrained bulk modulus and thermal expansion coefficient of HD C−S−H should be used. The following relations are used to calculate these parameters.

$$K_{u,HD}^{hom} = K_{HD}^{hom} + \frac{(b_{HD}^{hom})^2}{\frac{1}{N_{HD}^{hom}} + \frac{\phi_{HD}}{K_f}} \quad ; \quad \alpha_{u,HD}^{hom} = \alpha_{d,HD}^{hom} + \frac{Q^{hom} + \phi_{HD}\alpha_{f,HD} - b_{HD}^{hom}\alpha_{d,HD}^{hom}}{\frac{K_{HD}^{hom}}{b_{HD}^{hom}}\left(\frac{1}{N_{HD}^{hom}} + \frac{\phi_{HD}}{K_f}\right) + b_{HD}^{hom}} \tag{22}$$

where $\alpha_{f,HD}$ is the thermal expansion coefficient of the pore fluid in HD C-S-H. Considering the anomalous thermal expansion of confined fluids, two different thermal expansion coefficients are considered for the pore fluid in HD C-S-H and the one in LD C-S-H and macro pores, denoted by $\alpha_{f,LDV}$. We assume the same pore fluid compression modulus $K_f$ for all parts of the microstructure.

### 5.1.2. Cement paste

The microstructure of the cement paste for second homogenization step is constituted by five main phases: HD and LD C−S−H, Portlandite, anhydrous clinker and the macro-porosity. These phases are respectively represented by HD, LD, CH, CK and V subscripts. Among these phases, Portlandite crystals and anhydrous clinker can be considered as non-porous solids, while LD and HD C−S−H are porous solids. However, we assumed that in the poromechanics tests the porosity in HD C−S−H was not active and so the only porous phase in the microstructure is the LD C−S−H. Consequently the homogenization of the poroelastic properties at this step should be performed using the framework presented in section 3.2.1 for multiphase porous materials. The homogenized drained bulk modulus and shear modulus can be evaluated from equation (7):

$$K_{CP}^{hom} = f_{LD} K_{LD}^{hom} A_{LD,CP}^v + f_{HD} K_{u,HD}^{hom} A_{HD,CP}^v + f_{CH} k_{CH} A_{CH,CP}^v + f_{CK} k_{CK} A_{CK,CP}^v \tag{23}$$

$$G_{CP}^{hom} = f_{LD} G_{LD}^{hom} A_{LD,CP}^d + f_{HD} G_{HD}^{hom} A_{HD,CP}^d + f_{CH} g_{CH} A_{CH,CP}^d + f_{CK} g_{CK} A_{CK,CP}^d \tag{24}$$

Assuming spherical shapes for solids and pores, the strain localization tensor parameters, $A_{r,CP}^v$ and $A_{r,CP}^d$, of each phase can be estimated using equation (A.10) considering four solid phases and a pore volume, $(r = \text{HD, LD, CH, CK}, \phi)$. For HD C–S–H the homogenized undrained bulk modulus $k_{u,\text{HD}}^{\text{hom}}$ should be used. Considering the self-consistent scheme we should take $k_0 = K_{\text{CP}}^{\text{hom}}$, $g_0 = G_{\text{CP}}^{\text{hom}}$ and the parameters $\alpha_0$ and $\beta_0$ are given by following relations:

$$\alpha_0 = \frac{3K_{\text{CP}}^{\text{hom}}}{3K_{\text{CP}}^{\text{hom}} + 4G_{\text{CP}}^{\text{hom}}} \quad ; \quad \beta_0 = \frac{6\left(K_{\text{CP}}^{\text{hom}} + 2G_{\text{CP}}^{\text{hom}}\right)}{5\left(3K_{\text{CP}}^{\text{hom}} + 4G_{\text{CP}}^{\text{hom}}\right)} \tag{25}$$

The homogenized thermal parameter $\kappa$ can be obtained from equation (11):

$$\kappa_{\text{CP}}^{\text{hom}} = f_{\text{LD}}\alpha_{d,\text{LD}}^{\text{hom}} K_{\text{LD}}^{\text{hom}} A_{\text{LD,CP}}^v + f_{\text{HD}}\alpha_{u,\text{HD}}^{\text{hom}} K_{u,\text{HD}}^{\text{hom}} A_{\text{HD,CP}}^v \\ + f_{\text{CH}}\alpha_{\text{CH}} k_{\text{CH}} A_{\text{CH,CP}}^v + f_{\text{CK}}\alpha_{\text{CK}} k_{\text{CK}} A_{\text{CK,CP}}^v \tag{26}$$

The drained thermal expansion coefficient is then calculated as $\alpha_{d,\text{CP}}^{\text{hom}} = \kappa_{\text{CP}}^{\text{hom}}/K_{\text{CP}}^{\text{hom}}$. The homogenized thermal parameter $Q$ can be evaluated from equation (12):

$$Q_{\text{CP}}^{\text{hom}} = f_{\text{LD}}\left(\alpha_s\left(1-A_{\text{LD,CP}}^v\right)\left(1-b_{\text{LD}}^{\text{hom}}\right) + Q_{\text{LD}}^{\text{hom}}\right) + f_{\text{HD}}\alpha_{u,\text{HD}}^{\text{hom}}\left(1-A_{\text{HD,CP}}^v\right) \\ + f_{\text{CH}}\alpha_{\text{CH}}\left(1-A_{\text{CH,CP}}^v\right) + f_{\text{CK}}k_{\text{CK}}\left(1-A_{\text{CK,CP}}^v\right) \tag{27}$$

The active porosity is calculated as $\phi_{\text{CP}}^{\text{act}} = f_{\text{LD}}\phi_{\text{LD}} + f_{\text{V}}$. The expressions of the other homogenized poroelastic parameters $b_{\text{CP}}^{\text{hom}}$ and $N_{\text{CP}}^{\text{hom}}$, presented in [7], can be obtained respectively from equations (14) and (15). By evaluation of the homogenized parameters and the active porosity, the remaining parameters ($K_u$, $K_s$, $K_\phi$, $B$, $\Lambda$, $\alpha_u$, $\alpha_\phi$) can be evaluated. The homogenized thermal pressurization coefficient $\Lambda^{\text{hom}}$ which is used for the model calibration is calculated as:

$$\Lambda^{\text{hom}} = \frac{Q^{\text{hom}} + \phi^{\text{act}}\alpha_{f,\text{LDV}} - b^{\text{hom}}\alpha_d^{\text{hom}}}{\dfrac{1}{N^{\text{hom}}} + \dfrac{\phi^{\text{act}}}{K_f} + \dfrac{\left(b^{\text{hom}}\right)^2}{K_d^{\text{hom}}}} \tag{28}$$

## 5.2. Model calibration

The model parameters are summarized in Table (1). The properties of the microstructure phases are the same as the ones used by Ulm *et al*. [8]. The volumetric thermal expansion coefficient of Portlandite $\alpha_{\text{CH}}$ is taken equal to $7.0\times10^{-5}\,(°\text{C})^{-1}$ [40]. Note that assuming the complete hydration of the cement paste ($f_{\text{CK}}=0$), the properties of the clinker phase are not required. The principal unknown parameters are $k_s$, $g_s$, $\alpha_s$, $\alpha_{f,\text{HD}}$ and $\alpha_{f,\text{LDV}}$. The calibration of these parameters is done by minimizing the error between the experimentally evaluated parameters and the results of the homogenization model using a least-squares method. From the calibration of the mechanical properties in [7] we have $k_s = 25.0$ GPa and $g_s = 18.4$ GPa. The known macroscopic parameters are $\alpha_d^{\text{exp}} = 6\times10^{-5}\,(°\text{C})^{-1}$ and $\Lambda^{\text{exp}}$ given by equation (17). The calibration procedure is done in three steps. Note that $\alpha_{f,\text{HD}}$ and $\alpha_{f,\text{LDV}}$ are different and can vary with temperature but $\alpha_s$ is assumed to be constant. In section 2.2 we have seen that the thermal expansion of confined fluids increases with decreasing pore size. Knowing that the size of pores in HD C-S-H is smaller than the pores in LD C-S-H [41][42], it seems that $\alpha_{f,\text{HD}}$ should be greater than $\alpha_{f,\text{LDV}}$. The first calibration step is performed at 50°C assuming that at this temperature $\alpha_{f,\text{HD}} = \alpha_{f,\text{LDV}}$, i.e. the pore fluid thermal expansion is no more influenced by confinement in nanoscale pores. Based on the discussion in section 2.2, this assumption is compatible with the negligible pressure dependency of water thermal expansion at this temperature and also with the experimental results of Xu *et al.* [15][16]. The calibration of $\alpha_s$ and $\alpha_{f,\text{HD}} = \alpha_{f,\text{LDV}}$ at 50°C is done by minimizing the error defined below:

$$E_r = \left( \frac{\alpha_d^{\exp} - \alpha_d^{\text{hom}}}{\alpha_d^{\exp}} \right)^2 + \left( \frac{\Lambda^{\exp} - \Lambda^{\text{hom}}}{\Lambda^{\exp}} \right)^2 \qquad (29)$$

The calibration is done using a computer program which calculates the homogenized properties and the error for different combinations of $\alpha_s$ and $\alpha_{f,\text{HD}} = \alpha_{f,\text{LDV}}$. The minimum error is found for $\alpha_s = 4.2 \times 10^{-5} \, (\text{°C})^{-1}$ and $\alpha_{f,\text{HD}} = \alpha_{f,\text{LDV}} = 4.8 \times 10^{-4} \, (\text{°C})^{-1}$ (Figure (2-a)). The calibrated pore fluid thermal expansion coefficient at 50°C is very close to the one of pure bulk water at the same temperature that is $4.57 \times 10^{-4} \, (\text{°C})^{-1}$. This is compatible with the assumption made above concerning the negligible effect of the confinement in nanoscale pores at 50°C. The small difference of about 5% between the two thermal expansion coefficients can be attributed to the effect of dissolved ions.

The second step concerns the calibration of $\alpha_{f,\text{HD}}$. Knowing the value of $\alpha_s$ which is assumed to be constant, the only unknown variable for evaluation of $\alpha_d^{\text{hom}}$ at each temperature is $\alpha_{f,\text{HD}}$. This parameter is needed for evaluation of $\alpha_{u,\text{HD}}^{\text{hom}}$ from equation (22). Neglecting the variations of $\alpha_d^{\exp}$ with temperature between 25 and 50°C, $\alpha_{f,\text{HD}}$ should also remain constant equal to $4.8 \times 10^{-4} \, (\text{°C})^{-1}$.

The last calibration step is performed for $\alpha_{f,\text{LDV}}$. Knowing $\alpha_s$ and $\alpha_{f,\text{HD}}$ from the previous steps, $\alpha_{f,\text{LDV}}$ for different temperatures can be evaluated by minimizing the error between $\Lambda^{\text{hom}}$ from equation (28) and $\Lambda^{\exp}$ from equation (17). The variations of the error between $\Lambda^{\text{hom}}$ and $\Lambda^{\exp}$ for three different temperatures are presented in Figure (2-b).

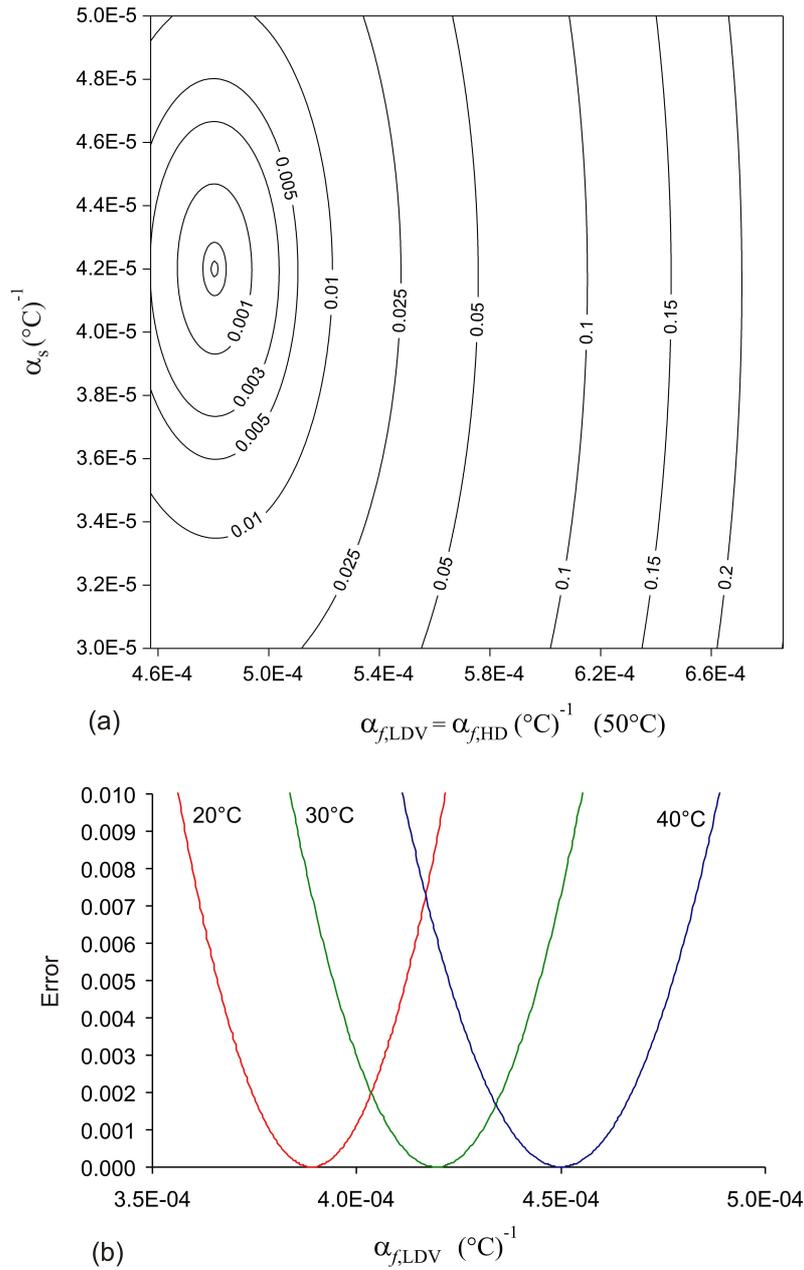

**Figure (2): (a) Contour plot of calculated error for different values of thermal expansion coefficients $\alpha_s$ and $\alpha_{f,\text{LDV}} = \alpha_{f,\text{HD}}$ at 50°C. The minimum error is equal to $1.88 \times 10^{-7}$ for $\alpha_s = 4.2 \times 10^{-5} \, (\text{°C})^{-1}$ and $\alpha_{f,\text{HD}} = \alpha_{f,\text{LDV}} = 4.8 \times 10^{-4} \, (°C)^{-1}$. (b) Variations of error for evaluation of $\alpha_{f,\text{LDV}}$ for different temperatures.**

The calculated $\alpha_{f,\text{HD}}$ and $\alpha_{f,\text{LDV}}$ are presented in Figure (3) and compared with the thermal expansion coefficients of pure bulk water and 0.5 mol/l NaOH bulk solution. We can see that $\alpha_{f,\text{HD}}$ and $\alpha_{f,\text{LDV}}$ are greater than the thermal expansions of bulk fluids. This anomalous thermal expansion of the cement paste pore fluid is due to the confinement in the nanoscale pores and the presence of dissolved ions. The effect of confinement can be analysed by calculating the confined

/bulk ratio of thermal expansion. This ratio reflects the effect of pore structure on the thermal expansion and is almost independent of presence of dissolved ions in the case of cement pore fluid which is mainly composed of univalent ions [1][17]. The confined/bulk ratio for the cement paste pore fluid can be calculated by assuming that the thermal expansion of the bulk fluid in the cement paste is equal to the thermal expansion of 0.5mol/l NaOH solution [1]. The results are presented in Figure (4) and compared with the confined/bulk ratio for pure water confined in 5.0 nm and 7.4 nm pores of silica glass [15][16]. We can see that the confined/bulk ratio for HD C-S-H is greater than the one of 5.0 nm pores, meaning that the average pore size in HD C-S-H should be smaller than 5.0 nm. This is compatible with the estimation of pore sizes in HD C-S-H from the results of mercury intrusion porosimetry. The mercury intrusion porosimetry does not basically permit to differentiate between LD and HD C-S-H porosity. However, a simple evaluation may be done assuming that the mercury can not access HD C-S-H porosity. This is similar to the assumption made by Tennis and Jennings [43] for evaluation of the volume fraction of LD C−S−H by analysing the results of surface area measurements by nitrogen sorption. These authors assumed that none of the pores in HD C−S−H are accessible to nitrogen. The empirical relation obtained using this assumption for volume fraction of LD C−S−H is widely used in the literature. Assuming that HD C-S-H is not accessible to mercury, from the results of this test as presented in section 4, it seems that HD C-S-H is composed of pores smaller than 6 nm. The confined/bulk ratio for LD C-S-H is smaller than the one of 5.0 nm pores and is closer to the one of 7.4 nm pores.

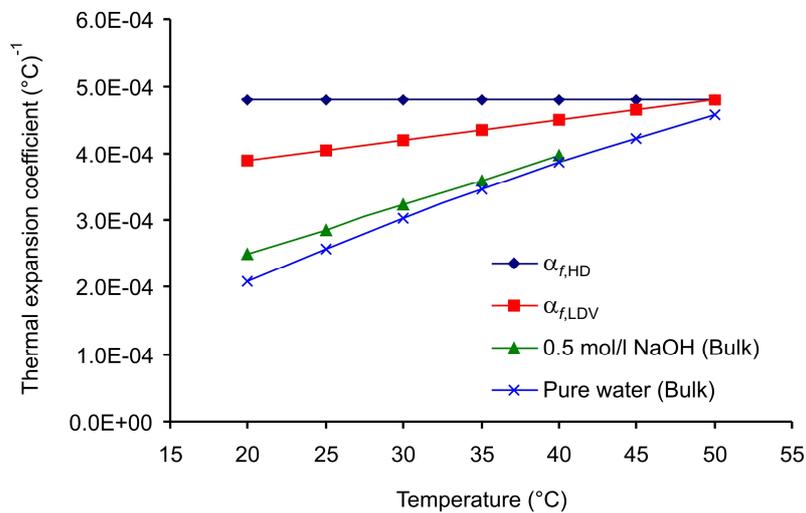

**Figure (3): Calibration of thermal expansion of cement paste pore fluid. Comparison with the thermal expansions of pure bulk water and 0.5 mol/l NaOH bulk solution**

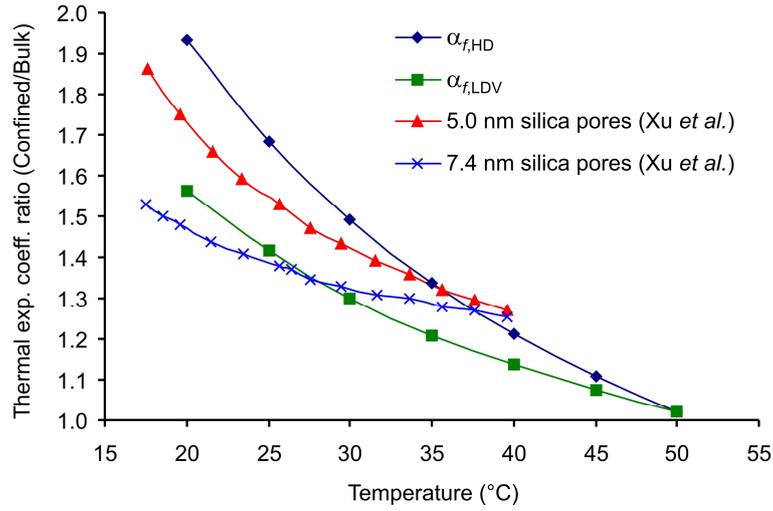

**Figure (4): Thermal expansion anomaly for cement paste pore fluid. Comparison with experimental results of Xu et al. [15][16] for thermal expansion anomaly of pure water confined in the pores of silica glass**

By calibration of necessary parameters, now we can proceed with extrapolation of the experimentally evaluated thermal expansion and thermal pressurization parameters to cement pastes with different water-to-cement ratios.

## *5.3. Effect of water-to-cement ratio*

The lower w/c limit is chosen equal to 0.4 as this is approximately the lowest w/c for which a complete hydration can be obtained. Figure (5) presents the variations of $\alpha_d$ and $\alpha_\phi$ with w/c. The pore volume thermal expansion $\alpha_\phi$ is calculated as $\alpha_\phi^{\text{hom}} = \left(b^{\text{hom}}\alpha_d^{\text{hom}} - Q^{\text{hom}}\right)/\phi^{\text{act}}$. Like for the modulus $K_\phi$ presented in [7], the possibility of evaluation of $\alpha_\phi$ is a considerable advantage of the presented association between the results of macroscopic experimental study and the homogenization method, as experimental evaluation of these parameters is very difficult. We can observe the decrease of $\alpha_d$ and more significant decrease of $\alpha_\phi$ with w/c. This decrease of thermal expansion coefficients by increasing w/c is mainly due to the decrease of Portlandite volume fraction in the microstructure, as presented in Figure (6). Note that Portlandite thermal expansion $\alpha_{\text{CH}} = 7.0 \times 10^{-5} \left(°\text{C}\right)^{-1}$, evaluated in [40] using time-of-flight neutron diffraction, is greater than the calibrated thermal expansion of C-S-H solid $\alpha_s = 4.2 \times 10^{-5} \left(°\text{C}\right)^{-1}$. Consequently a reduction of the Portlandite fraction results in a decrease of the homogenized thermal expansion coefficients. Figure (6) shows also the increase of active porosity with w/c increase. This is mainly due to increase of the quantity of remaining water after the complete hydration of cement clinker. The available water is completely consumed in hydration reactions for w/c close to 0.4. For higher w/c a quantity of water is not consumed and forms the macro-porosity of the cement paste. For w/c lower than 0.56, the evaluated $\alpha_\phi$ is greater than $\alpha_d$ but becomes smaller than $\alpha_d$ for higher w/c. The difference between these parameters is smaller than 11% and shows that the assumption $\alpha_\phi = \alpha_d$ made in [1] for back analysis of the undrained heating test is acceptable. The induced error on evaluation of the thermal pressurization coefficient $\Lambda$ assuming $\alpha_\phi = \alpha_d$ is smaller than 1%. This is mainly due to significant difference between the thermal expansion of water and the ones of the material, $\alpha_d$ and $\alpha_\phi$. The assumption $\alpha_\phi = \alpha_d$ is commonly made in the literature due to the difficulty of experimental evaluation of $\alpha_\phi$. It should be mentioned that the coefficients $\alpha_d$ and $\alpha_\phi$ in Figure (5) do not vary with temperature, because the calibrated pore fluid thermal expansion for HD C-S-H is constant. This is due to the measured $\alpha_d$ which shows negligible temperature dependency.

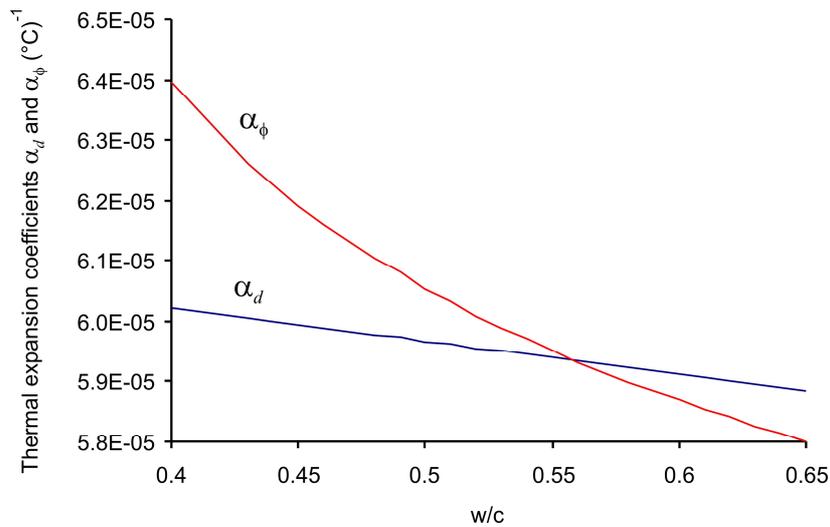

**Figure (5): Effect of water-to-cement ratio on drained thermal expansion $\alpha_d$ and pore volume thermal expansion $\alpha_\phi$**

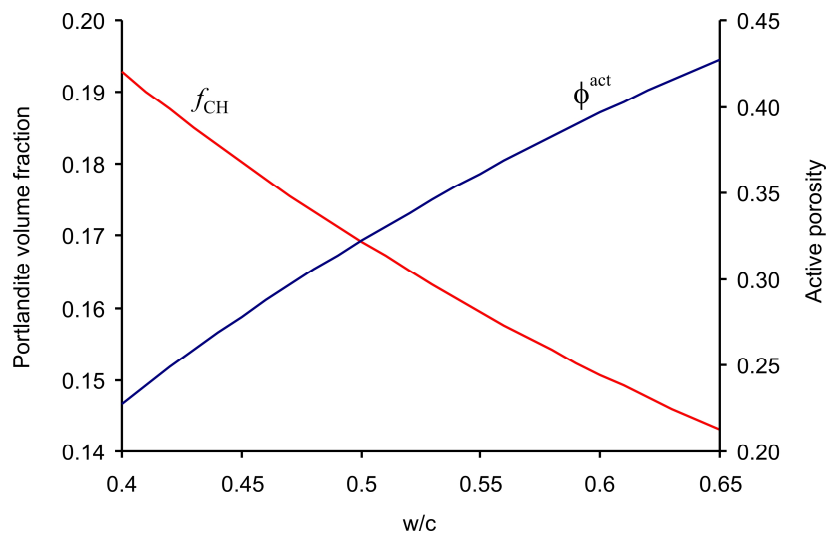

**Figure (6): Variations of Portlandite volume fraction and active porosity with water-to-cement ratio**

Figure (7) shows the variations of undrained thermal expansion coefficient $\alpha_u$ and the thermal pressurization coefficient $\Lambda$ with w/c for two different temperatures, 25°C and 40°C. Both parameters increase with temperature due to the increase of thermal expansion coefficient of the pore fluid. A higher w/c results in a higher $\Lambda$ and a lower $\alpha_u$, mainly due to the significant reduction of $K_d$ and the resulting increase of $b$ with w/c increase. $K_d$ reduces from 9.81 GPa at w/c=0.4 to 3.97 GPa at w/c=0.65 and results in an increase of $b$ from 0.54 to 0.81 [7]. The significant increases of the terms $b^2/K_d$ in equation (28) and $b/K_d$ in equation (6) cause the reduction of $\Lambda$ and increase of $\alpha_u$ with w/c increase. The reduction of $K_d$ with w/c increase is mainly due to the increase of porosity with w/c, as presented in Figure (6). It would be ideal to compare these predictions of thermal expansion and thermal pressurization parameters with experimentally evaluated values, but unfortunately such experimental results for different w/c are not currently available.

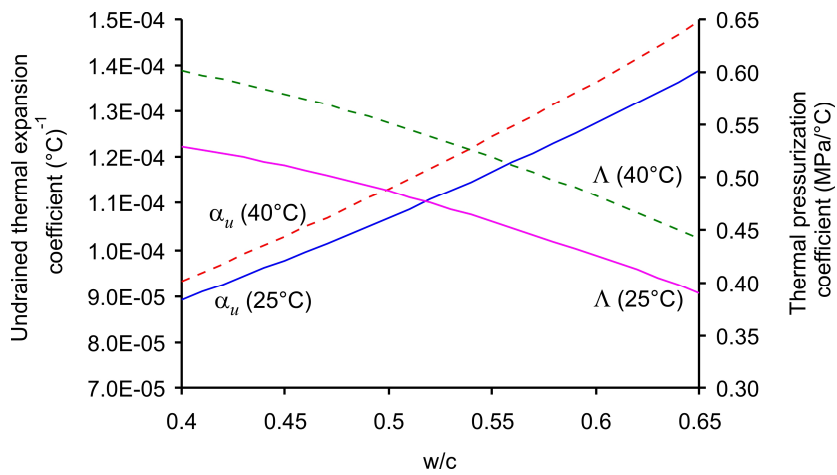

**Figure (7): Effect of water-to-cement ratio on undrained thermal expansion $\alpha_u$ and thermal pressurization coefficient $\Lambda$**

## 6. Conclusions

This paper is presented in the continuity of the approach introduced in [7] for association of the results of a macro-scale experimental study with the micromechanics modelling and homogenization technique. This approach is applied here to the results of drained and undrained heating tests performed on a hardened class G cement paste with w/c=0.44 [1]. The main purpose is to extrapolate the experimentally evaluated thermal expansion and thermal pressurization parameters to cement pastes with different water-to-cement ratios. The used multi-scale model is capable of predicting the macroscopic thermo-poroelastic parameters of a hardened cement paste by knowing the volume fractions and the thermo-elastic properties of the constituents of its microstructure. The model calibration for thermal parameters is performed by means of some simplification assumptions, but revealed interesting information about the anomalous thermal expansion behaviour of cement paste pore fluid. The calculated thermal expansion anomaly for the pore fluid in HD C-S-H and also in LD C-S-H and macro-porosity show a good compatibility with the experimental results of Xu *et al*. [15][16][17] on thermal expansion of confined fluids. The calibrated model permits to calculate the thermal expansion and thermal pressurization parameters for cement pastes with different w/c. Moreover, it permits to evaluate the pore volume thermal expansion coefficient $\alpha_\phi$ which is very difficult to measure experimentally. This capacity of parameter prediction for different conditions and the better understanding of the results of the tests, as also presented in [7], clearly demonstrate the advantages of the association of macroscopic laboratory experiments and micromechanics modelling. This approach reduces significantly the number of laboratory tests needed to characterize the complete set of thermo-poroelastic parameters of a cement paste. This is a great advantage for experimental studies, as due to the very low permeability of the material, the laboratory tests are usually long.

## 7. Acknowledgment

The author wishes to thank Professor Jean Sulem and Manh-Huyen Vu for fruitful discussions and comments.

# 8. Appendix A: Homogenization of thermo-poroelastic parameters

This appendix is dedicated to derivation of the equations of micro-thermo-poroelasticity and homogenization method. The aim of classical homogenization techniques is to replace an actual heterogeneous complex body by a fictitious homogeneous one which behaves globally in the same way. Continuum micromechanics is mainly concerned with statically homogeneous materials for which it is possible to define a representative elementary volume (REV). Over the REV, the average values of local stress and strain fields in the actual heterogeneous body are equal to the macroscopic values of stress and strain fields derived by solving the boundary value problem of a homogeneous body constituted by this fictitious homogeneous material [31]. This requires that, for the mechanical behaviour under investigation, the characteristic length $d$ of the considered heterogeneity and deformation mechanism to be much smaller than the size $l$ of the studied volume element. Moreover, $l$ must be sufficiently smaller than the characteristic dimension $L$ of the whole body.

After the scale separation, the three steps of homogenization method as mentioned by Zaoui [31] are: *description* (or representation), *concentration* (or localization) and *homogenization* (or upscaling). The *description* step deals with identification of different "mechanical" phases of the microstructure in the REV of the considered heterogeneous material, and both geometrical and mechanical characteristics of these phases. A phase, in the sense of continuum micromechanics, is a material domain that can be identified, at a given scale, with on-average constant material properties. The *concentration* step is concerned with the mechanical modelling of the interactions between the phases and the link between the local stress and strain fields within the REV and the macroscopic quantities of stress and strain. The last step deals with the *homogenization* of the macroscopic properties by combining the local constitutive equations, averaging the stresses and the strains over the REV and the concentration relations. Homogenization delivers estimated values of macroscopic poroelastic properties of the REV as a function of the geometrical and mechanical properties of different phases of the microstructure of the material.

## *8.1. Representation*

The volume $V_0$ of the REV of a heterogeneous material is composed of $n$ different phases with volumes $V_r$, $r = 1 \ldots n$, and volume fractions denoted by $f_r = V_r/V_0$. We consider that there is only one porous phase with volume $V_\phi$ and porosity $\phi = V_\phi/V_0$. The number of solid phases is therefore $m = n - 1$ with total volume $V_s$. The tensor of elastic moduli of each phase is denoted by $\mathbf{c}_r$. In the case of isotropy of the solid phases, the tensor of elastic moduli can be written as the sum of a volumetric and a deviatoric part:

$$\mathbf{c}_r = 3k_r \mathbf{J} + 2g_r \mathbf{K} \qquad (A.1)$$

where $k_r$ and $g_r$ are the bulk modulus and shear modulus of the phase $r$ respectively. $\mathbf{J}_{ijkl} = 1/3 \delta_{ij}\delta_{kl}$ is the volumetric part of the fourth-order symmetric unit tensor $\mathbf{I}$ and $\mathbf{K} = \mathbf{I} - \mathbf{J}$ is the deviatoric part. $\mathbf{I}$ is defined as $\mathbf{I}_{ijkl} = 1/2\left(\delta_{ik}\delta_{jl} + \delta_{il}\delta_{jk}\right)$ and $\delta_{ij}$ stands for Kronecker delta. The tensor of thermal expansion coefficients of each phase is denoted by $\boldsymbol{\alpha}_r$ which is reduced to $\alpha_r \mathbf{1}$ in the isotropic case with $\mathbf{1}_{ij} = \delta_{ij}$.

## *8.2. Concentration*

The concentration problem is presented by assuming homogeneous boundary conditions on the REV [36][37]. *Homogeneous stress boundary conditions* correspond to prescribed surface tractions $\underline{T}$ on the boundary $\partial V$ of the REV:

$$\text{on } \partial V: \quad \underline{T} = \boldsymbol{\Sigma} \cdot \underline{n} \tag{A.2}$$

where $\boldsymbol{\Sigma}$ is the macroscopic stress tensor and $\underline{n}$ is the unit outward normal at the boundary. From (A.2) it can be shown that the macroscopic stress $\boldsymbol{\Sigma}$ is equal to the volume average of the microscopic equilibrated (i.e., divergence free) stress field $\boldsymbol{\sigma}(\underline{x})$ in the REV [31].

$$\boldsymbol{\Sigma} = \langle \boldsymbol{\sigma} \rangle_V \tag{A.3}$$

where $\langle z \rangle_V = (1/V) \int_V z(\underline{x}) \mathrm{d}V$ stands for the volume average of quantity $z$ over domain $V$. Similarly, *homogeneous strain boundary conditions* are associated to prescribed displacements $\underline{u}$ at the boundary:

$$\text{on } \partial V: \quad \underline{u} = \mathbf{E} \cdot \underline{x} \tag{A.4}$$

where $\underline{x}$ is the microscopic position vector and $\mathbf{E}$ is the macroscopic strain tensor which is equal to the volume average of the microscopic compatible (i.e., derived from a displacement field) strain field $\boldsymbol{\varepsilon}(\underline{x})$ in the REV [31].

$$\mathbf{E} = \langle \boldsymbol{\varepsilon} \rangle_V \tag{A.5}$$

For the homogeneous boundary conditions (A.2) or (A.4) Hill's lemma is presented in the following form [30][31]:

$$\langle \boldsymbol{\sigma} : \boldsymbol{\varepsilon} \rangle = \langle \boldsymbol{\sigma} \rangle : \langle \boldsymbol{\varepsilon} \rangle = \boldsymbol{\Sigma} : \mathbf{E} \tag{A.6}$$

The Hill lemma is relevant for any stress and strain compatible with either a homogeneous stress boundary condition (A.2) or a homogeneous strain boundary condition (A.4), irrespective of a link between $\boldsymbol{\sigma}$ and $\boldsymbol{\varepsilon}$ through a constitutive law.

In the framework of linear elasticity, the local strain and stress fields, $\boldsymbol{\varepsilon}(\underline{x})$ and $\boldsymbol{\sigma}(\underline{x})$, are related to macroscopic strain and stress, $\mathbf{E}$ and $\boldsymbol{\Sigma}$, through fourth-order localization tensors $\mathbf{A}(\underline{x})$ and $\mathbf{B}(\underline{x})$ respectively:

$$\boldsymbol{\varepsilon}(\underline{x}) = \mathbf{A}(\underline{x}) : \mathbf{E} \quad ; \quad \boldsymbol{\sigma}(\underline{x}) = \mathbf{B}(\underline{x}) : \boldsymbol{\Sigma} \tag{A.7}$$

By inserting equation (A.7) in equations (A.5) and (A.3), the following relations for localizations tensors are obtained:

$$\langle \mathbf{A} \rangle_V = \mathbf{I} \quad ; \quad \langle \mathbf{B} \rangle_V = \mathbf{I} \tag{A.8}$$

For a heterogeneous material composed of homogeneous phases, a linear phase strain localization tensor can be introduced [8]:

$$\langle \mathbf{\varepsilon} \rangle_{V_r} = \langle \mathbf{A} \rangle_{V_r} : \mathbf{E} \quad ; \quad \sum_{r=1}^{n} f_r \langle \mathbf{A} \rangle_{V_r} = \mathbf{I} \tag{A.9}$$

In the isotropic case $\langle \mathbf{A} \rangle_{V_r}$ is reduced to $\langle \mathbf{A} \rangle_{V_r} = A_r^v \mathbf{J} + A_r^d \mathbf{K}$ where $A_r^v$ and $A_r^d$ are volumetric and deviatoric strain localization coefficients. For an Eshelbian type morphology [38], i.e. an spherical inclusion embedded in a reference medium, an estimate of the strain localization tensor of phase $r$, assuming the isotropy of the local and the reference medium is given by [31]:

$$A_r^v = \frac{\left(1+\alpha_0 \left(k_r/k_0 - 1\right)\right)^{-1}}{\sum_r f_r \left(1+\alpha_0 \left(k_r/k_0 - 1\right)\right)^{-1}} \quad ; \quad A_r^d = \frac{\left(1+\beta_0 \left(g_r/g_0 - 1\right)\right)^{-1}}{\sum_r f_r \left(1+\beta_0 \left(g_r/g_0 - 1\right)\right)^{-1}} \tag{A.10}$$

With

$$\alpha_0 = \frac{3k_0}{3k_0 + 4g_0} \quad ; \quad \beta_0 = \frac{6(k_0 + 2g_0)}{5(3k_0 + 4g_0)} \tag{A.11}$$

where $k_0$ and $g_0$ are bulk modulus and shear modulus of the reference medium. According to the choice of the reference medium in these equations one can distinguish two different homogenization schemes: the *Mori-Tanaka* scheme [44] in which the reference medium is chosen to be the matrix phase; the *Self-consistent* scheme [45] in which the reference medium is the homogenized medium. The Mori-Tanaka scheme is mostly adapted to the composite materials in which the continuous matrix plays a prominent morphological role in the behaviour of the material. In this case, $k_0$ and $g_0$ are taken equal to the elastic parameters of the material phase which is considered as the reference medium. The Self consistent scheme is adequate for materials, such as polycrystals, whose phases are dispersed in the RVE so that none of them plays any specific morphological role [31]. In the case of self-consistent scheme, $k_0$ and $g_0$ are taken equal to the homogenized elastic properties which are unknown in advance. This point is further explained in section 8.3.1.

## *8.3. Homogenization*

The equations of micro-poroelasticity and the homogenization of the poroelastic properties can be derived on a REV submitted to a homogeneous strain boundary condition and two eigenstresses $\mathbf{\sigma}^p$ and $\mathbf{\sigma}^T$ corresponding respectively to application of a pore pressure and a temperature variation. The constitutive relation in the microstructure is given in the form:

$$\boldsymbol{\sigma}(\underline{x}) = \mathbf{c}(\underline{x}) : \boldsymbol{\varepsilon}(\underline{x}) + \boldsymbol{\sigma}^p(\underline{x}) + \boldsymbol{\sigma}^T(\underline{x}) \tag{A.12}$$

where $\mathbf{c}(\underline{x})$ is the tensor of local elastic moduli which is equal to zero in the pore volume:

$$\mathbf{c}(\underline{x}) = \begin{cases} \mathbf{c}_r(\underline{x}) & \text{in } V_s \\ 0 & \text{in } V_\phi \end{cases} \tag{A.13}$$

$\boldsymbol{\sigma}^p(\underline{x})$ is an eigenstress applied to the pore volume of the material:

$$\boldsymbol{\sigma}^p(\underline{x}) = \begin{cases} 0 & \text{in } V_s \\ p\mathbf{1} & \text{in } V_\phi \end{cases} \tag{A.14}$$

$\boldsymbol{\sigma}^T(\underline{x})$ is an eigenstress applied to the solid phase of the material:

$$\boldsymbol{\sigma}^T(\underline{x}) = \begin{cases} \boldsymbol{\kappa}(\underline{x})T & \text{in } V_s \\ 0 & \text{in } V_\phi \end{cases} \tag{A.15}$$

where $\boldsymbol{\kappa} = \mathbf{c} : \boldsymbol{\alpha}$. The linear elastic nature of the microscopic behaviour allows decomposing the problem in three sub-problems. In the first sub-problem the eigenstresses $\boldsymbol{\sigma}^p$ and $\boldsymbol{\sigma}^T$ are equal to zero and the REV is submitted only to the homogeneous strain boundary conditions. The displacement on the boundary of the REV is equal to zero in the second and third sub-problems and the system is subjected to the eigenstresses $\boldsymbol{\sigma}^p$ and $\boldsymbol{\sigma}^T$ respectively. These three sub-problems are denoted respectively by $()'$, $()''$ and $()'''$ superscripts.

### 8.3.1. Sub-problem 1

In the first sub-problem the eigenstresses are equal to zero and the local strain is given using equation (A.7):

$$\boldsymbol{\sigma}'(\underline{x}) = \mathbf{c}(\underline{x}) : \boldsymbol{\varepsilon}'(\underline{x}) \quad ; \quad \boldsymbol{\varepsilon}'(\underline{x}) = \mathbf{A}(\underline{x}) : \mathbf{E} \tag{A.16}$$

The macroscopic stress $\boldsymbol{\Sigma}'$ is equal to the volume average of the microscopic equilibrated stress field $\boldsymbol{\Sigma}' = \langle \boldsymbol{\sigma}' \rangle_V$. By inserting equation (A.16) in this relation the following expression is obtained where $\mathbf{C}^{\text{hom}}$ can be viewed as the tensor of the overall *effective* moduli of the heterogeneous porous material.

$$\boldsymbol{\Sigma}' = \mathbf{C}^{\text{hom}} : \mathbf{E} \quad ; \quad \mathbf{C}^{\text{hom}} = \langle \mathbf{c} : \mathbf{A} \rangle_V = \sum_{r=1}^{n} f_r \mathbf{c}_r : \langle \mathbf{A} \rangle_{V_r} \tag{A.17}$$

In the isotropic case $\mathbf{C}^{\text{hom}}$ can be presented in the following form:

$$\mathbf{C}^{\text{hom}} = 3K_d^{\text{hom}} \mathbf{J} + 2G^{\text{hom}} \mathbf{K} \tag{A.18}$$

The homogenized bulk and shear moduli in the isotropic case are calculated using following relations:

$$K_d^{\text{hom}} = \sum_{r=1}^{n} f_r k_r A_r^v \quad ; \quad G^{\text{hom}} = \sum_{r=1}^{n} f_r g_r A_r^d \tag{A.19}$$

In the case of an Eshelbian type morphology, the strain concentration coefficients $A_r^v$ and $A_r^d$ can be evaluated using equation (A.10). The self-consistent homogenization scheme is associated with the choice of $k_0 = K_d^{\text{hom}}$ and $g_0 = G^{\text{hom}}$ which are unknown in advance. Consequently, equation (A.19) can not be solved directly and the homogenized elastic properties should be calculated using iterative calculations.

The porosity change $d\phi = \phi - \phi_0$ can be calculated from the average volumetric strain in the pore volume, $(d\phi)' = -\phi_0 \mathbf{1} : \langle \boldsymbol{\varepsilon}' \rangle_{V_\phi}$. By using equation (A.16) in this equation we can identify the tensor of *effective* Biot's coefficients $\mathbf{b}^{\text{hom}}$:

$$(d\phi)' = -\mathbf{b}^{\text{hom}} : \mathbf{E} \quad ; \quad \mathbf{b}^{\text{hom}} = \phi_0 \mathbf{1} : \langle \mathbf{A} \rangle_{V_\phi} = \mathbf{1} : \left( \mathbf{I} - \sum_{r=1}^{m} f_r \langle \mathbf{A} \rangle_{V_r} \right) \quad (A.20)$$

For the isotropic case, Biot's coefficient is obtained from the following simplified equation:

$$b^{\text{hom}} = 1 - \sum_{r=1}^{m} f_r A_r^v \quad (A.21)$$

The average local stress in the solid phase is obtained using equation (A.16) equal to $\langle \boldsymbol{\sigma}' \rangle_{V_s} = \langle \mathbf{c} : \boldsymbol{\varepsilon}' \rangle_{V_s} = \langle \mathbf{c} : \mathbf{A} \rangle_{V_s} : \mathbf{E}$. From equation (A.9) the average local strain in the solid phase is equal to $\langle \boldsymbol{\varepsilon}' \rangle_{V_s} = \langle \mathbf{A} \rangle_{V_s} : \mathbf{E}$. Using these equations the relation between the average local stress and strains in the solid phase permits to evaluate the tensor of effective solid moduli $\mathbf{C}_s^{\text{hom}}$:

$$\langle \boldsymbol{\sigma}' \rangle_{V_s} = \mathbf{C}_s^{\text{hom}} : \langle \boldsymbol{\varepsilon}' \rangle_{V_s} \quad ; \quad \mathbf{C}_s^{\text{hom}} = \langle \mathbf{c} : \mathbf{A} \rangle_{V_s} : \langle \mathbf{A} \rangle_{V_s}^{-1} \quad (A.22)$$

In the isotropic case the expression of effective unjacketed modulus is written as:

$$K_s^{\text{hom}} = \sum_{i=1}^{m} f_r k_r A_r^v \bigg/ \sum_{i=1}^{m} f_r A_r^v \quad (A.23)$$

From equations (A.17) and (A.13) we have $\mathbf{C}^{\text{hom}} = (1 - \phi_0) \langle \mathbf{c} : \mathbf{A} \rangle_{V_s}$. Moreover, from equation (A.8) we have $\phi_0 \langle \mathbf{A} \rangle_{V_\phi} = \mathbf{I} - (1 - \phi_0) \langle \mathbf{A} \rangle_{V_s}$. Inserting these relations in equation (A.22) and using equation (A.20) we can obtain the following expression for the tensor of effective Biot's coefficients:

$$\mathbf{b}^{\text{hom}} = \mathbf{1} : \left( \mathbf{I} - \mathbf{C}^{\text{hom}} : \left( \mathbf{C}_s^{\text{hom}} \right)^{-1} \right) \quad (A.24)$$

In the isotropic case this relation is reduced to the well-known relation $b^{\text{hom}} = 1 - K_d^{\text{hom}} / K_s^{\text{hom}}$.

### 8.3.2. Sub-problem 2

In the second sub-problem the displacement on the boundary of the REV is equal to zero ($\mathbf{E}'' = \langle \boldsymbol{\varepsilon}'' \rangle_V = 0$) and the system is subjected to the eigenstress $\boldsymbol{\sigma}^p$ defined in equation (A.14). The local stress tensor is given by:

$$\boldsymbol{\sigma}''(\underline{x}) = \mathbf{c}(\underline{x}) : \boldsymbol{\varepsilon}''(\underline{x}) + \boldsymbol{\sigma}^p(\underline{x}) \quad (A.25)$$

The macroscopic stress tensor for this sub-problem is calculated by application of Hill's lemma, equation (A.6), on the stress field of second sub-problem $\boldsymbol{\sigma}''$ and the strain field of first sub-problem $\boldsymbol{\varepsilon}'$:

$$\mathbf{E}:\boldsymbol{\Sigma}'' = \langle \boldsymbol{\varepsilon}':\boldsymbol{\sigma}'' \rangle_V = \langle \boldsymbol{\varepsilon}':\mathbf{c}:\boldsymbol{\varepsilon}'' \rangle_V + \langle \boldsymbol{\varepsilon}':\boldsymbol{\sigma}^p \rangle_V \tag{A.26}$$

Using equation (A.16) and then, by application of Hill's lemma the first term in the right-hand side of equation (A.26) is found to be equal to zero:

$$\langle \boldsymbol{\varepsilon}':\mathbf{c}:\boldsymbol{\varepsilon}'' \rangle_V = \langle \boldsymbol{\sigma}':\boldsymbol{\varepsilon}'' \rangle_V = \boldsymbol{\Sigma}':\langle \boldsymbol{\varepsilon}'' \rangle_V = 0 \tag{A.27}$$

Now by introducing equation (A.16) in the second term of the right-hand side of equation (A.26) and then, using equation (A.14) one finds the following equations which permit to identify the tensor of effective Biot's coefficients, as presented in equation (A.20):

$$\boldsymbol{\Sigma}'' = \langle \boldsymbol{\sigma}^p : \mathbf{A} \rangle_V = \phi_0 \mathbf{1}:\langle \mathbf{A} \rangle_{V_\phi} p = p\mathbf{b}^{\text{hom}} \tag{A.28}$$

Knowing that $\mathbf{E}'' = \langle \boldsymbol{\varepsilon}'' \rangle_V = 0$, the variation of the porosity is given by following expression:

$$(d\phi)'' = -\phi_0 \mathbf{1}:\langle \boldsymbol{\varepsilon}'' \rangle_{V_\phi} = (1-\phi_0)\mathbf{1}:\langle \boldsymbol{\varepsilon}'' \rangle_{V_s} = \mathbf{1}:\sum_{r=1}^{m} f_r \mathbf{c}_r^{-1}:\langle \boldsymbol{\sigma}'' \rangle_{V_r} \tag{A.29}$$

The average local stress in the REV is given by

$$\langle \boldsymbol{\sigma}'' \rangle_V = \phi_0 \langle \boldsymbol{\sigma}'' \rangle_{V_\phi} + \sum_{r=1}^{m} f_r \langle \boldsymbol{\sigma}'' \rangle_{V_r} \quad ; \quad \langle \boldsymbol{\sigma}'' \rangle_V = \boldsymbol{\Sigma}'' \tag{A.30}$$

Using equations (A.20), (A.28) and (A.30) and knowing that $\langle \boldsymbol{\sigma}'' \rangle_{V_\phi} = p\mathbf{1}$ we find:

$$\sum_{r=1}^{m} f_r \langle \boldsymbol{\sigma}'' \rangle_{V_r} = p\left(\mathbf{b}^{\text{hom}} - \phi_0 \mathbf{1}\right) = p\mathbf{1}:\sum_{r=1}^{m} f_r \left(\mathbf{I} - \langle \mathbf{A} \rangle_{V_r}\right) \tag{A.31}$$

From this equation we can obtain the average strain over a phase of the solid volume [8]:

$$\langle \boldsymbol{\sigma}'' \rangle_{V_r} = p\mathbf{1}:\left(\mathbf{I} - \langle \mathbf{A} \rangle_{V_r}\right) \qquad \text{in } V_s \tag{A.32}$$

Inserting equation (A.32) in equation (A.29) we obtain the following expression for the effective Biot skeleton modulus $N^{\text{hom}}$:

$$(d\phi)'' = \frac{p}{N^{\text{hom}}} \quad ; \quad \frac{1}{N^{\text{hom}}} = \mathbf{1}:\sum_{r=1}^{m} f_r \mathbf{c}_r^{-1}:\left(\mathbf{1} - \mathbf{1}:\langle \mathbf{A} \rangle_{V_r}\right) \tag{A.33}$$

In the isotropic case this relation is reduced to:

$$\frac{1}{N^{\text{hom}}} = \sum_{r=1}^{m} \frac{f_r \left(1 - A_r^v\right)}{k_r} \tag{A.34}$$

### 8.3.3. Sub-problem 3

In the third sub-problem the displacement on the boundary of the REV is equal to zero ($\mathbf{E}''' = \langle \boldsymbol{\varepsilon}''' \rangle_V = 0$) and the system is subjected to the eigenstress $\boldsymbol{\sigma}^T$ defined in equation (A.15). The local stress tensor is given by:

$$\boldsymbol{\sigma}'''(\underline{x}) = \mathbf{c}(\underline{x}) : \boldsymbol{\varepsilon}'''(\underline{x}) + \boldsymbol{\sigma}^T(\underline{x}) \tag{A.35}$$

The macroscopic stress tensor for this sub-problem is calculated by application of Hill's lemma, equation (A.6), on the stress field of the third sub-problem $\boldsymbol{\sigma}'''$ and the strain field of first sub-problem $\boldsymbol{\varepsilon}'$:

$$\mathbf{E} : \boldsymbol{\Sigma}''' = \langle \boldsymbol{\varepsilon}' : \boldsymbol{\sigma}''' \rangle_V = \langle \boldsymbol{\varepsilon}' : \mathbf{c} : \boldsymbol{\varepsilon}''' \rangle_V + \langle \boldsymbol{\varepsilon}' : \boldsymbol{\sigma}^T \rangle_V \tag{A.36}$$

Using equation (A.16) and then, by application of Hill's lemma the first term in the right-hand side of equation (A.26) is found to be equal to zero:

$$\langle \boldsymbol{\varepsilon}' : \mathbf{c} : \boldsymbol{\varepsilon}''' \rangle_V = \langle \boldsymbol{\sigma}' : \boldsymbol{\varepsilon}''' \rangle_V = \boldsymbol{\Sigma}' : \langle \boldsymbol{\varepsilon}''' \rangle_V = 0 \tag{A.37}$$

By introducing equation (A.16) in the second term of the right-hand side of equation (A.36) and then, using equation (A.15) one finds the following equations which permit to identify the tensor of effective coefficients $\boldsymbol{\kappa}^{\text{hom}}$:

$$\boldsymbol{\Sigma}''' = \langle \boldsymbol{\sigma}^T : \mathbf{A} \rangle_V = \boldsymbol{\kappa}^{\text{hom}} T \quad ; \quad \boldsymbol{\kappa}^{\text{hom}} = \langle \boldsymbol{\kappa} : \mathbf{A} \rangle_V = \sum_{r=1}^n f_r \boldsymbol{\kappa}_r : \langle \mathbf{A} \rangle_{V_r} \tag{A.38}$$

Knowing that $\mathbf{E}''' = \langle \boldsymbol{\varepsilon}''' \rangle_V = 0$, the variation of the porosity is given by following expression:

$$(d\phi)''' = -\phi_0 \mathbf{1} : \langle \boldsymbol{\varepsilon}''' \rangle_{V_\phi} = \mathbf{1} : \sum_{r=1}^m f_r \langle \boldsymbol{\varepsilon}''' \rangle_{V_r} \tag{A.39}$$

Using equations (A.35) and (A.15) the average local stress in the REV is given by:

$$\langle \boldsymbol{\sigma}''' \rangle_V = \sum_{r=1}^m f_r \mathbf{c}_r : \langle \boldsymbol{\varepsilon}''' \rangle_{V_r} + T \sum_{r=1}^m f_r \boldsymbol{\kappa}_r \quad ; \quad \langle \boldsymbol{\sigma}''' \rangle_V = \boldsymbol{\Sigma}''' \tag{A.40}$$

Using equation (A.38) in equation (A.40) we obtain:

$$\sum_{r=1}^m f_r \mathbf{c}_r : \langle \boldsymbol{\varepsilon}''' \rangle_{V_r} = -T \sum_{r=1}^m f_r \boldsymbol{\kappa}_r : \left( \mathbf{I} - \langle \mathbf{A} \rangle_{V_r} \right) \tag{A.41}$$

Knowing that $\boldsymbol{\alpha} = \mathbf{c}^{-1} : \boldsymbol{\kappa}$ we can evaluate the average strain in a phase of the solid volume:

$$\langle \boldsymbol{\varepsilon}''' \rangle_{V_r} = -\boldsymbol{\alpha}_r : \left( \mathbf{I} - \langle \mathbf{A} \rangle_{V_r} \right) T \quad \text{in } V_s \tag{A.42}$$

Inserting equation (A.42) in equation (A.39) we obtain the following expression for the effective coefficient $Q^{\text{hom}}$:

$$(d\phi)''' = -Q^{\text{hom}} T \quad ; \quad Q^{\text{hom}} = \mathbf{1} : \sum_{r=1}^m f_r \boldsymbol{\alpha}_r : \left( \mathbf{I} - \langle \mathbf{A} \rangle_{V_r} \right) \tag{A.43}$$

In the isotropic case this relation is reduced to:

$$Q^{\text{hom}} = \sum_{r=1}^{m} f_r \alpha_r \left(1 - A_r^v\right) \quad \text{(A.44)}$$

### 8.3.4. Macroscopic equations

The summation of the macroscopic stresses and porosity variations of the three sub-problems permit to retrieve the equations of macro-poroelasticity. From equations (A.17), (A.28) and (A.38) we have:

$$\mathbf{\Sigma} = \mathbf{\Sigma}' + \mathbf{\Sigma}'' + \mathbf{\Sigma}''' = \mathbf{C}^{\text{hom}} : \mathbf{E} + \mathbf{b}^{\text{hom}} p + \mathbf{\kappa}^{\text{hom}} T \quad \text{(A.45)}$$

Similarly, from equations (A.20), (A.33) and (A.43) we find:

$$d\phi = (d\phi)' + (d\phi)'' + (d\phi)''' = -\mathbf{b}^{\text{hom}} : \mathbf{E} + \frac{p}{N^{\text{hom}}} - Q^{\text{hom}} T \quad \text{(A.46)}$$

## *8.4. Multi-scale porous material*

A particular situation, mentioned by Ulm *et al.* [8] and Dormieux *et al.* [35], which is not addressed directly in the standard micro-poroelasticity as presented in the previous section is the case of a porous material in which the pore volume manifests itself at two or several different scales. These two pore volumes are connected and there is one homogeneous pore pressure in all parts of the pore volume. The homogenization of poroelastic properties of such a porous material needs a multi-step homogenization technique. The first step of this procedure is the homogenization of the porous phases which have the smallest-length scales porosity. This step is performed using the standard homogenization equations as presented in the previous section. The next step of homogenization procedure is concerned with a heterogeneous material composed of some porous phases, some solid phases and a pore volume with a greater length scale than the one inside the porous phases. Let us consider a two-scale porous material in which the pore volume exhibits two different scales I and II ($V_\phi = V_\phi^{\text{I}} + V_\phi^{\text{II}}$), i.e. a micro-porosity and a macro-porosity. This material is composed of $l$ ($l \leq m$) porous phases with the porosities $\phi_r^{\text{I}}$, $m - l$ solid phases and a pore volume with the porosity $\phi^{\text{II}}$. The total porosity of the material is therefore given by:

$$\phi = \sum_{r=1}^{l} f_r \phi_r^{\text{I}} + \phi^{\text{II}} \quad \text{(A.47)}$$

Standard homogenization permits to evaluate the poroelastic properties of the $l$ porous phases of level I ($\mathbf{c}_r^{\text{I}}$, $\mathbf{b}_r^{\text{I}}$, $N_r^{\text{I}}$). Referring to the sub-problems defined in the previous section, using equations (A.47) and (A.46), the variation of the porosity in the first sub-problem is given by:

$$(d\phi)' = \sum_{r=1}^{l} f_r (d\phi_r^{\text{I}})' + (d\phi^{\text{II}})' = -\sum_{r=1}^{l} f_r \mathbf{b}_r^{\text{I}} : \langle \mathbf{\varepsilon}' \rangle_{V_r} - \phi_0^{\text{II}} \mathbf{1} : \langle \mathbf{\varepsilon}' \rangle_{V_\phi^{\text{II}}} \quad \text{(A.48)}$$

Using equations (A.9) and knowing that $\mathbf{b}_r^{\text{I}} = 0$ for the non-porous phases, equation (A.48) can be re-written as:

$$(d\phi)' = -\sum_{r=1}^{m} f_r \mathbf{b}_r^I : \langle \mathbf{A} \rangle_{V_r} : \mathbf{E} - \left( \mathbf{I} - \sum_{r=1}^{m} f_r \langle \mathbf{A} \rangle_{V_r} \right) : \mathbf{E} \tag{A.49}$$

Consequently the homogenized tensor of Biot effective stress coefficients is found to be:

$$(d\phi)' = -\mathbf{b}^{\text{hom}} : \mathbf{E} \quad ; \quad \mathbf{b}^{\text{hom}} = \mathbf{1} - \sum_{r=1}^{m} \left( f_r \langle \mathbf{A} \rangle_{V_r} : (\mathbf{1} - \mathbf{b}_r^I) \right) \tag{A.50}$$

It can be verified that equation (A.50) is reduced to equation (A.20) when all solid phases are non-porous ($\mathbf{b}_r^I = 0$), i.e. the porosity is taking effect in a single length scale. For the isotropic case equation (A.50) is presented in the following simplified form:

$$b^{\text{hom}} = 1 - \sum_{r=1}^{m} \left( f_r A_r^v (1 - b_r^I) \right) \tag{A.51}$$

The variation of the porosity for the second sub-problem can be obtained using equations (A.47) and (A.46):

$$(d\phi)'' = \sum_{r=1}^{l} f_r \left( d\phi_r^I \right)'' + \left( d\phi^{II} \right)'' = \sum_{r=1}^{l} f_r \left( -\mathbf{b}_r^I : \langle \boldsymbol{\varepsilon}'' \rangle_{V_r} + \frac{p}{N_r^I} \right) - \phi_0^{II} \mathbf{1} : \langle \boldsymbol{\varepsilon}'' \rangle_{V_\phi^{II}} \tag{A.52}$$

Using the right-hand side equality of equation (A.29) in equation (A.52) and noting that $\mathbf{b}_r^I = 0$ and $1/N_r^I = 0$ in the non-porous phases, the following expression can be obtained:

$$(d\phi)'' = \sum_{r=1}^{m} f_r \left( (\mathbf{1} - \mathbf{b}_r^I) : \langle \boldsymbol{\varepsilon}'' \rangle_{V_r} + \frac{p}{N_r^I} \right) \tag{A.53}$$

From equation (A.45) for the first homogenization step of each solid phase we have:

$$\langle \boldsymbol{\varepsilon}'' \rangle_{V_r} = \left( \mathbf{c}_r^I \right)^{-1} : \left( \langle \boldsymbol{\sigma}'' \rangle_{V_r} - p \mathbf{b}_r^I \right) \quad \text{in } V_s \tag{A.54}$$

The average local stress in the phase $r$ of the solid phase can be evaluated using equations (A.28) and (A.30) and $\mathbf{b}^{\text{hom}}$ of multiscale porous material from equation (A.50):

$$\langle \boldsymbol{\sigma}'' \rangle_{V_r} = p \left( \mathbf{1} - \mathbf{1} : \langle \mathbf{A} \rangle_{V_r} + \mathbf{b}_r^I : \langle \mathbf{A} \rangle_{V_r} \right) \quad \text{in } V_s \tag{A.55}$$

This relation is equivalent to equation (A.32) for simple porous materials. Replacing equations (A.55) and (A.54) in (A.53) and using the relation $(\mathbf{1} - \mathbf{b}_r^I) : (\mathbf{c}_r^I)^{-1} = \mathbf{1} : (\mathbf{c}_{sr}^I)^{-1}$ from (A.24) we obtain:

$$(d\phi)'' = \frac{p}{N^{\text{hom}}} \quad ; \quad \frac{1}{N^{\text{hom}}} = \sum_{r=1}^{m} f_r \left( \mathbf{1} : (\mathbf{c}_{sr}^I)^{-1} : (\mathbf{I} - \langle \mathbf{A} \rangle_{V_r}) : (\mathbf{1} - \mathbf{b}_r^I) + \frac{1}{N_r^I} \right) \tag{A.56}$$

When all solid phases are non-porous ($\mathbf{b}_r^I = 0$, $1/N_r^I = 0$, $\mathbf{c}_{sr}^I = \mathbf{c}_r$), equation (A.56) is reduced to (A.33). For the isotropic case (A.56) can be re-written as:

$$\frac{1}{N^{\text{hom}}} = \sum_{r=1}^{m} f_r \left( \frac{(1 - A_r^v)(1 - b_r^I)}{k_{sr}^I} + \frac{1}{N_r^I} \right) \tag{A.57}$$

The porosity variation for the third sub-problem is obtained using (A.47) and (A.46):

$$\left(d\phi\right)''' = \sum_{r=1}^{l} f_r \left(d\phi_r^{\mathrm{I}}\right)''' + \left(d\phi^{\mathrm{II}}\right)''' = \sum_{r=1}^{l} f_r \left(-\mathbf{b}_r^{\mathrm{I}} : \langle \boldsymbol{\varepsilon}''' \rangle_{V_r} - Q_r^{\mathrm{I}} T\right) - \phi_0^{\mathrm{II}} \mathbf{1} : \langle \boldsymbol{\varepsilon}''' \rangle_{V_\phi^{\mathrm{II}}} \quad (A.58)$$

Using the right-hand side equality of (A.39) in (A.58) and noting that $\mathbf{b}_r^{\mathrm{I}} = 0$ and $Q_r^{\mathrm{I}} = 0$ in the non-porous phases, the following expression can be obtained:

$$\left(d\phi\right)''' = \sum_{r=1}^{m} f_r \left(\left(\mathbf{1} - \mathbf{b}_r^{\mathrm{I}}\right) : \langle \boldsymbol{\varepsilon}''' \rangle_{V_r} - Q_r^{\mathrm{I}} T\right) \quad (A.59)$$

Using (A.45) for the first homogenization step of each solid phase and equations (A.38) and (A.40) the average local strain in phase *r* of the solid phase is evaluated as:

$$\langle \boldsymbol{\varepsilon}''' \rangle_{V_r} = -\boldsymbol{\alpha}_r^{\mathrm{I}} : \left(\mathbf{I} - \langle \mathbf{A} \rangle_{V_r}\right) T \qquad \text{in } V_s \quad (A.60)$$

This relation is equivalent to (A.42) for simple porous materials. Replacing (A.60) in (A.59) we obtain:

$$\left(d\phi\right)''' = -Q^{\mathrm{hom}} T \quad ; \quad Q^{\mathrm{hom}} = \sum_{r=1}^{m} f_r \left(\boldsymbol{\alpha}_r^{\mathrm{I}} : \left(\mathbf{I} - \langle \mathbf{A} \rangle_{V_r}\right) : \left(\mathbf{1} - \mathbf{b}_r^{\mathrm{I}}\right) + Q_r^{\mathrm{I}}\right) \quad (A.61)$$

When all solid phases are non-porous ($\mathbf{b}_r^{\mathrm{I}} = 0$, $Q_r^{\mathrm{I}} = 0$), equation (A.61) is reduced to equation (A.43).

# 10. List of symbols

| | |
|---|---|
| **A** | Strain localization tensor |
| $A^v$ | Volumetric strain localization coefficient |
| $A^d$ | Deviatoric strain localization coefficient |
| $b$ | Biot's effective stress coefficient |
| **b** | Tensor of Biot's effective stress coefficients |
| $B$ | Skempton's coefficient |
| **c** | Microscopic tensor of elastic moduli |
| $\mathbf{c}_s$ | Microscopic tensor of solid moduli |
| **C** | Macroscopic tensor of elastic moduli |
| $\mathbf{C}_s$ | Macroscopic tensor of solid moduli |
| **E** | Macroscopic strain tensor |
| $E$ | Macroscopic volumetric strain |
| $f$ | Volume fraction of microstructure phase |
| $g$ | Microscopic shear modulus |
| $G$ | Shear modulus |
| $k$ | Microscopic bulk modulus |
| $K_d$ | Drained bulk modulus |
| $K_p$ | Drained pore volume modulus |
| $K_s$ | Unjacketed modulus |
| $K_\phi$ | Unjacketed pore volume modulus |
| $K_u$ | Undrained bulk modulus |
| $K_f$ | Pore fluid bulk modulus |
| $m$ | Number of solid phases |
| $n$ | Number of phases |
| $N$ | Biot's skeleton modulus |
| $p$ | Pore pressure |
| $Q$ | Thermal porosity change coefficient |
| $T$ | Temperature |
| $V$ | Total volume |
| $V_\phi$ | Pore volume |
| $V_s$ | Solid volume |
| $\underline{x}$ | Position vector |
| $\alpha_d$ | Drained thermal expansion coefficient |
| $\alpha_\phi$ | Pore volume thermal expansion coefficient |
| $\alpha_u$ | Undrained thermal expansion coefficient |
| $\alpha_f$ | Pore fluid thermal expansion coefficient |
| $\boldsymbol{\alpha}$ | Tensor of thermal expansion coefficients |
| $\boldsymbol{\varepsilon}$ | Microscopic strain tensor |
| $\phi$ | Lagrangian porosity |
| $\phi^{act}$ | Active porosity |
| $\kappa$ | Thermal stress coefficient |
| $\Lambda$ | Thermal pressurization coefficient |
| $\boldsymbol{\sigma}$ | Microscopic stress tensor |
| $\boldsymbol{\sigma}^P$ | Eigenstress related to pore pressure |
| $\boldsymbol{\sigma}^T$ | Eigenstress related to temperature |
| $\Sigma$ | Macroscopic mean stress |
| $\Sigma_d$ | Macroscopic Terzaghi effective stress |
| $\boldsymbol{\Sigma}$ | Macroscopic stress tensor |
| CH | Subscript for Portlandite |
| CK | Subscript for cement clinker |
| CP | Subscript for cement paste |
| HD | Subscript for high density C-S-H |
| LD | Subscript for low density C-S-H |
| CSH | Subscript for C-S-H |
| s | Subscript for C-S-H solid |
| V | Subscript for macro porosity |
| hom | Superscript for homogenized parameter |
| exp | Superscript for experimentally evaluated parameter |